\documentclass[journal]{IEEEtran}
\IEEEoverridecommandlockouts

\usepackage{cite}
\usepackage{amsmath,amssymb,amsfonts,bm}
\usepackage{algorithmic}
\usepackage{algorithm}
\usepackage{graphicx}
\usepackage{textcomp}
\usepackage[table]{xcolor}
\usepackage{xcolor}
\usepackage{wrapfig}
\usepackage{multirow}
\usepackage{tabularx}
\usepackage{url}
\usepackage{import}    
\usepackage{comment}  
\usepackage{dirtytalk}  
\usepackage{stfloats}
\usepackage{soul}
\usepackage{adjustbox}
\usepackage{xfrac}
\usepackage{makecell}
\usepackage{epsfig}
\usepackage{epsf}
\usepackage[hidelinks]{hyperref}
\usepackage{mdframed}
\usepackage{subcaption} 

\usepackage[most]{tcolorbox}

\usepackage{amsthm}  

\newtheorem{lemma}{Lemma}

\definecolor{mygray}{RGB}{240,240,240} 
\definecolor{lblue}{RGB}{237, 247, 255} 
\definecolor{lred}{RGB}{248, 232, 232} 

\definecolor{lblue}{RGB}{237, 247, 255} 
\definecolor{lblue}{RGB}{237, 247, 255} 

\definecolor{lightbluebox}{HTML}{D6D6FF}
\definecolor{lightpurplebox}{HTML}{F0C8F0}

\definecolor{titlepurple}{HTML}{421A36}     
\definecolor{lightpurple}{HTML}{F7F2F5}     

\def\BibTeX{{\rm B\kern-.05em{\sc i\kern-.025em b}\kern-.08em
    T\kern-.1667em\lower.7ex\hbox{E}\kern-.125emX}}

\usepackage{etoolbox}
\usepackage{tikz}

\newrobustcmd*{\fillsquare}[1]{\tikz{\filldraw[draw=#1,fill=#1] (0,0)
rectangle (0.25cm,0.25cm);}}

\newrobustcmd*{\mycircle}[1]{\tikz{\filldraw[draw=#1,fill=#1] (0,0) circle [radius=0.1cm];}}

\newrobustcmd*{\mytriangle}[1]{\tikz{\filldraw[draw=#1,fill=#1] (0.3cm,0.3cm) -- (0cm,0.3cm) -- (0.15cm,0cm);}}

\newrobustcmd*{\fillnabla}[1]{\tikz{\filldraw[draw=#1,fill=#1] (0.2cm,0) -- (0.1cm,0.2cm) -- (0,0);}}

\usepackage{comment}

\specialcomment{rimuovere}{}{}
\specialcomment{riscrivere}{\begingroup\color{orange}}{\endgroup}
\specialcomment{commentolungo}{\begingroup\color{gray}}{\endgroup}
\specialcomment{new}{\begingroup\color{blue}}{\endgroup}
\specialcomment{warning}{\begingroup\color{red}}{\endgroup}

\def\BibTeX{{\rm B\kern-.05em{\sc i\kern-.025em b}\kern-.08em
    T\kern-.1667em\lower.7ex\hbox{E}\kern-.125emX}}
\begin{document}

\title{Holographic \& Channel-Aware \\ Distributed Detection of a Non-cooperative Target}

\author{Domenico Ciuonzo,~\IEEEmembership{Senior Member,~IEEE}, Alessio Zappone,~\IEEEmembership{Fellow,~IEEE} \\ Marco Di Renzo,~\IEEEmembership{Fellow,~IEEE} and Ciro d'Elia

\thanks{Manuscript received 15th Jun. 2025; revised 20th Nov 25; accepted 25th Jan 26.}
\thanks{D. Ciuonzo is with the Dept. of Electrical Engineering and Information Technologies (DIETI) at Univ. of Naples Federico II, Italy (domenico.ciuonzo@unina.it).}

\thanks{A. Zappone and C. D'Elia are with Dept. of Electrical and Information Engineering ``Maurizio Scarano'', Univ. of Cassino and Southern Lazio, Italy. A. Zappone is also with Consorzio Nazionale Interuniversitario per le Telecomunicazioni (CNIT), Italy (alessio.zappone@unicas.it; delia@unicas.it).}

\thanks{M. Di Renzo is with Universit\'e Paris-Saclay, CNRS, CentraleSup\'elec, Laboratoire des Signaux et Syst\`emes, 3 Rue Joliot-Curie, 91192 Gif-sur-Yvette, France. (marco.di-renzo@universite-paris-saclay.fr), and with King's College London, Centre for Telecommunications Research -- Department of Engineering, WC2R 2LS London, United Kingdom (marco.di\_renzo@kcl.ac.uk).}

\thanks{Research leading to these results has received funding from Project ``Garden'', CUP H53D23000480001 funded by EU in NextGeneration EU plan, Mission 4 Component 1, through the Italian ``Bando Prin 2022 - D.D. 104 del 02-02-2022`` by MUR.
This manuscript reflects only the authors’ views and opinions and the Ministry cannot be considered responsible for them.
\\
The work of A. Zappone has received funding from the Horizon Europe project TWIN6G under grant agreement number 101182794.
\\
The work of M. Di Renzo was supported in part by the European Union through the Horizon Europe project COVER under grant agreement number 101086228, the Horizon Europe project UNITE under grant agreement number 101129618, the Horizon Europe project INSTINCT under grant agreement number 101139161, and the Horizon Europe project TWIN6G under grant agreement number 101182794, as well as by the Agence Nationale de la Recherche (ANR) through the France 2030 project ANR-PEPR Networks of the Future under grant agreement NF-PERSEUS 22-PEFT-004, and by the CHIST-ERA project PASSIONATE under grant agreements CHIST-ERA-22-WAI-04 and ANR-23-CHR4-0003-01.}}

\markboth{IEEE Transactions on Aerospace and Electronic Systems,~Vol.~*, No.~*, Month~2025}{Ciuonzo \MakeLowercase{\textit{et al.}}: Holographic \&
Channel-Aware Distributed Detection of a Non-cooperative Target}

\maketitle
\begin{abstract}
This work investigates Distributed Detection (DD) in Wireless Sensor Networks (WSNs), where spatially distributed sensors transmit binary decisions over a shared flat-fading channel. To enhance fusion efficiency, a reconfigurable metasurface is positioned in the near-field of a few receive antennas, enabling a holographic architecture that harnesses large-aperture gains with minimal RF hardware.
A generalized likelihood ratio test is derived for fixed metasurface settings, and two low-complexity joint design strategies are proposed to optimize both fusion and metasurface configuration. These suboptimal schemes achieve a balance between performance, complexity, and system knowledge.
The goal is to ensure reliable detection of a localized phenomenon at the fusion center, under energy-efficient constraints aligned with IoT requirements. Simulation results validate the effectiveness of the proposed holographic fusion, even under simplified designs.
\end{abstract}


\section{Introduction}
\label{Introduction}

\subsection{Motivation and Background}

\IEEEPARstart{D}{istributed} Detection (DD) \cite{viswanathan1997,blum2002distributed} plays a key role in cyberphysical systems, supporting large-scale sensing for monitoring, surveillance, and situational awareness. In this setting, spatially distributed wireless sensors
transmit local decisions--either binary or soft--to a Fusion Center (FC), which determines the presence of a target or Phenomenon of Interest (PoI). 
Achieving reliable inference in such systems requires scalable and energy-efficient communication architectures that operate under bandwidth constraints and leverage the underlying physics of wireless propagation.

Conventional orthogonal access schemes scale poorly in dense Wireless Sensor Networks (WSNs), motivating the use of Multiple Access Channels (MACs), where sensors transmit simultaneously over a shared medium~\cite{li2007distributed}. While spectrally-efficient, this approach introduces interference that degrades detection performance. To address this, distributed Multiple-Input Multiple-Output (MIMO) architectures have been proposed~\cite{zhang2008optimal,Ciuonzo2012}, in which a multi-antenna FC emulates a virtual uplink MIMO channel~\cite{jiang2007multiuser}.
This architecture provides receive diversity and robustness, significantly improving the fusion performance.

The advent of \emph{massive MIMO}~\cite{lu2014overview} extends these benefits by exploiting large antenna arrays to suppress fading and improve energy efficiency~\cite{ciuonzo2015,jiang2015massive}. However, fully-digital implementations require one Radio Frequency (RF) chain per antenna~\cite{molisch2017hybrid}, resulting in high complexity and power consumption. Furthermore, physical limitations such as mutual coupling and half-wavelength spacing constrain array size and scalability.

To address these limitations, recent efforts have explored hardware-efficient alternatives based on \emph{metasurfaces}--compact, sub-wavelength structures that can manipulate electromagnetic waves with fine spatial control~\cite{direnzo2020}. 
These architectures enable energy-efficient beamforming and spectrum shaping, and can be broadly categorized into: ($i$) Reconfigurable Intelligent Surfaces (RISs), which operate as smart reflectors in the propagation environment~\cite{huang2020holographic}, and ($ii$) Reconfigurable Holographic Surfaces (RHSs), which are embedded directly within transceiver architectures.
While RISs reshape the wireless environment to assist coverage and connectivity, RHSs are integrated into the physical layer of the receiver itself, enabling thin, low-power, and reconfigurable antenna arrays~\cite{jamali2020intelligent,interdonato2024approaching}. Their holographic nature inherently supports beamforming through wavefront synthesis, allowing precise radiation patterns with minimal RF hardware.

In this work, we focus on RHS-based receive architectures to support DD. 
This novel approach--holographic Decision Fusion (holographic DF)~\cite{Ciuonzo2025IoT}--aims to achieve high detection performance while significantly reducing hardware complexity and power consumption at the FC.

Despite their promise, \emph{existing RHS-based DD architectures assume idealized models}, such as perfectly known or spatially-homogeneous target statistics~\cite{Ciuonzo2025IoT}. However, real-world scenarios are often governed by \emph{spatial uncertainty}, where the statistical profile of the PoI varies across the sensing field in unknown and possibly nonlinear ways. 
This challenge invalidates conventional fusion rules--such as those based on the
likelihood ratio and its simplifications, e.g., \cite{Chen2004, Ciuonzo2012,ciuonzo2015}--and render static RHS configurations inefficient or even misleading.

This motivates the need for new inference frameworks and adaptive RHS design strategies that can \emph{learn and exploit spatially-varying statistics},  and do so under strict hardware and energy constraints. Addressing this challenge is the central contribution of this paper.

\subsection{Related Works}

\noindent
\textbf{DD in emerging wireless environments}: research on DD has progressed from early work on quantization and fusion over ideal links~\cite{chair1986,reibman1987} to modern studies that leverage WSNs and channel-aware rules~\cite{Chen2004}.
Recent interest has turned toward exploiting (virtual) MIMO \cite{zhang2008,Ciuonzo2012} and massive MIMO~\cite{ciuonzo2015,jiang2015massive,chawla2019,Chawla2021} front-ends to enhance detection performance and reduce energy expenditure. However, these works typically assume either idealized sensor models or known, fully-specified sensing channels. 

On the contrary, \emph{when uncertainty in the sensing process is considered} (e.g., target location uncertainty), \emph{the analysis is often limited to idealized settings}. These include, for instance, binary symmetric channel models~\cite{ciuonzo2017quantizer,cheng2019multibit,yang2023hybrid,mao2024multi}, which inherently decouple inference from communication by adhering to a suboptimal \emph{decode-then-fuse} approach~\cite{Ciuonzo2012}, as well as simplified wireless channel models, such as parallel-access channel models~\cite{Ciuonzo2025single}. 
In other words, DD strategies have either ($a$) handled complex wireless propagation \emph{without} sensor-level uncertainty, or ($b$) addressed sensing model imperfections \emph{without} accounting for realistic multi-access and MIMO effects.

\noindent
\textbf{Adoption of metasurfaces for distributed inference}: the use of programmable metasurfaces in distributed inference has \emph{predominantly focused on RISs}~\cite{fang2021,zhang2022worst,zhai2022beamforming,zhao2023ris}, aimed at enabling over-the-air computation (AirComp) by shaping the wireless medium.
Most existing works optimize RIS parameters to compute predefined nomographic functions under favorable assumptions (e.g., synchronized transmission, idealized measurement statistics, and real-time or long-term channel knowledge).
A smaller body of work has considered \emph{inference-oriented goals}, e.g., estimation~\cite{Ahmed2022,rajput2024joint} or detection~\cite{ge2024ris,mudkey2022wireless,Ciuonzo2025icassp}.
Nonetheless, \emph{even these contributions exhibit inherent limitations}: ($a$) they typically assume homogeneous or fully-specified sensing models, thereby neglecting the spatial heterogeneity and partial sensing information that characterize realistic deployments; ($b$) they primarily leverage metasurfaces for enhancing communication coverage, rather than exploiting their potential for holographic fusion systems.

\noindent
\textbf{Literature gap \& positioning}: In view of the aforementioned discussion, the role of RHSs in DD remains largely unexplored.
Recent efforts have begun to consider RHS-powered DD~\cite{Ciuonzo2025IoT}, but they stop short of addressing the \emph{joint challenge of sensing uncertainty and dynamically reconfigurable wireless environments}.
Consequently, the design solutions proposed in~\cite{Ciuonzo2025IoT} are \emph{of limited applicability in the present setup}: ($i$) they either presuppose a fully-specified and accurate sensing model (\say{FuC-i} design), which is often unavailable in practical scenarios, or ($ii$) they rely on overly simplistic assumptions (\say{IS} design) that fail to capture the inherent challenged of the considered DD problem.
While RISs have been extensively studied for enabling AirComp in WSNs, current works neither incorporate RHSs into the FC architecture nor adopt a goal-oriented design under practical complexity constraints.
In parallel, the design of holographic transceivers has gained momentum in wireless communications, yet their integration into inference-driven WSNs--where signal processing and propagation must be co-designed--remains an open problem.

\subsection{Summary of Contributions and Paper Organization}
Hence, the \textbf{main contributions} are summarized as follows: 
\begin{itemize}
\item We address the design of an RHS-assisted FC in DD systems, where the RHS matrix and the fusion rule are \emph{jointly optimized} in a \emph{goal-oriented} fashion. Unlike prior work, the considered model accounts for \emph{both} ($i$) unknown and spatially varying PoIs/targets and ($ii$) a composite channel model induced by the RHS and a reduced set of external feeds connected to a digital fusion processor.

\item To overcome the infeasibility of optimizing the RHS under a Generalized Likelihood Ratio (GLR) fusion rule, we adopt the \emph{deflection criterion} (in both standard and modified forms)~\cite{Picinbono1995,Quan2008} as a tractable surrogate and impose Widely-Linear (WL) fusion processing to avoid the exponential growth of GLR complexity with the number of sensors.
Within this framework, we introduce \emph{two complementary design objectives} that integrate environmental uncertainty directly into the physical-layer RHS configuration, thereby coupling sensing and fusion in a unified process.

\item Based on the aforementioned objective formulations, we formulate \emph{four optimization problems} for joint analog-digital processing.
The first two (\textbf{eFuC-0/1}) embed the expected detection probability of the unknown target into deflection-optimal WL-constrained optimization framework~\cite{Ciuonzo2025IoT}.
The second two (\textbf{bFuC-0/1}) couple analog-domain spatial filtering with digital-domain fusion using a WL filter-bank strategy~\cite{willsky2003generalized}, enabling implicit target localization and improved robustness under heterogeneous target conditions.

\item All formulations are solved via Alternating Optimization (AO) and Majorization–Minimization (MM)~\cite{sun2016majorization} methods, yielding closed-form iterative updates for the RHS matrix and WL fusion vector(s). 
A detailed per-step complexity analysis (Sec.~\ref{subsec:complexity}) is provided, highlighting suitability for WSNs with practical energy and latency constraints.

\item The proposed approaches are compared with: ($i$) the sensing-agnostic (joint) IS design in~\cite{Ciuonzo2025IoT}, ($ii$) a GLR-based fusion approach with no RHS optimization (or with IS-derived RHS), and ($iii$) fundamental upper bounds corresponding to ideal reporting channels.
\end{itemize}

The rest of the paper is structured as follows. Sec.~\ref{Model} details the system model under consideration.
Sec.~\ref{sec_fusion_rule_RIS} formulates the eFuC/bFuC design problem, whereas Sec.~\ref{sec:AO_based_solution} describes the proposed solution based on AO. Simulation-based performance evaluation is reported in Sec.~\ref{Sim_res}, while Sec.~\ref{Conclusions} ends the paper and outlines potential directions for future research.%
\footnote{\textbf{Notation} -- vectors (resp. matrices) are denoted with lower-case (resp. upper-case) bold letters; $\mathbb{E}\{\cdot\}$,
$\mathrm{var\{\cdot\}}$, $\mathrm{Cov(\cdot)}$, $\mathrm{PCov(\cdot)}$, $(\cdot)^{T}$, $(\cdot)^{\dagger}$, $\Re\left(\cdot\right)$, $\angle(\cdot)$ and $\left\Vert \cdot\right\Vert $ denote expectation,
variance, covariance, pseudocovariance, transpose, conjugate transpose, real part, phase, and Euclidean norm operators, respectively;
$j$ denotes the imaginary unit;
$\bm{O}_{N\times K}$
(resp. $\bm{I}_{N}$) denotes the $N\times K$ (resp. $N\times N$)
null (resp. identity) matrix; $\bm{0}_{N}$ (resp. $\bm{1}_{N}$)
denotes the null (resp. ones) vector of length $N$;
$\mathrm{diag}(\bm{a})$
denotes the diagonal matrix with $\bm{a}$ on the main diagonal; $\lambda_{max}(\bm{A})$ evaluates the highest eigenvalue from the positive-definite matrix $\bm{A}$; $\underline{\bm{a}}$
(resp. $\underline{\bm{A}}$) denotes the augmented vector (resp.
matrix) of $\bm{a}$ (resp. $\bm{A}$), that is $\underline{\bm{a}}\triangleq\left[\begin{array}{cc}
\bm{a}^{T} & \bm{a}^{\dagger}\end{array}\right]^{T}$ (resp. $\underline{\bm{A}}\triangleq\left[\begin{array}{cc}
\bm{A}^{T} & \bm{A}^{\dagger}\end{array}\right]^{T}$); $\Pr(\cdot)$ and $p(\cdot)$ denote probability mass functions (pmfs) and probability density functions (pdfs), while $\Pr(\cdot|\cdot)$ and $p(\cdot|\cdot)$ their corresponding conditional counterparts; $\mathcal{N}_{\mathbb{C}}(\bm{\mu},\bm{\Sigma})$ (resp. $\mathcal{N}(\bm{\mu},\bm{\Sigma})$) denotes a proper complex (resp. real) normal distribution with mean vector $\bm{\mu}$ and covariance matrix $\bm{\Sigma}$;
$\mathcal{Q}(\cdot)$ denotes the complementary cumulative
distribution function of a normal random variable
in its standard form, i.e. $\mathcal{N}(0,1)$;
$\mathcal{U}(a,b)$ denotes a uniform distribution with support $(a,b)$;
the symbols $\sim$ and $\propto$ mean \say{distributed as} and \say{proportional to}, respectively.}

\begin{figure}[ht]
    \includegraphics[width=1\columnwidth]{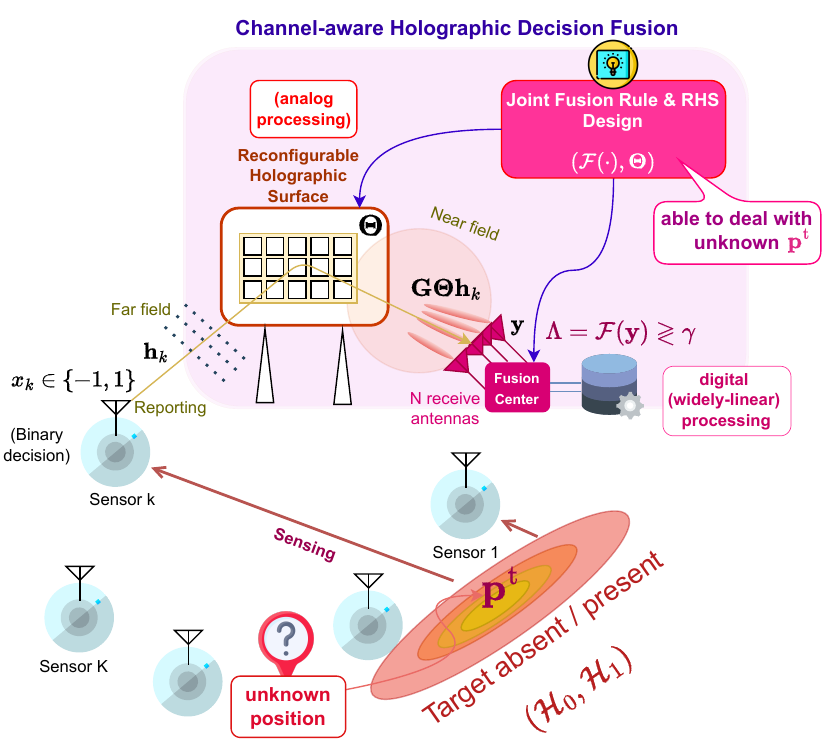}
	\caption{The RHS-assisted DF system model considered in the case of a non-cooperative target (\emph{unknown} location $\bm{p}^t$).}
 \label{fig:system.model}
\end{figure}

\section{System Model}\label{Model}
This section describes the DD problem in a WSN employing holographic DF (cf. Fig.~\ref{fig:system.model}). 
It outlines the target sensing and modulation assumptions (Sec.~\ref{subsec:sensing_model}), introduces the hybrid far-/near-field channel model (Sec.~\ref{subsec:channel_model}), and ends with the conditional statistical characterization of the received signal vector (Sec.~\ref{subsec:stat_char}).

\subsection{Sensing Model \& Modulation Format}\label{subsec:sensing_model}

We consider a network of $K$ low-cost sensors, indexed by $k\in\mathcal{K}\triangleq\{1,\ldots,K\}$, deployed across a given region of interest, whose objective is to collaboratively infer the presence ($\mathcal{H}_{1}$) or absence ($\mathcal{H}_{0}$) of target appearing in a surveillance area $\mathcal{A}$.
Under hypothesis $\mathcal{H}_{1}$, each sensor acquires a noisy realization of a spatially isotropic emission field, whose amplitude is subject to random fluctuations induced by the inherent stochasticity of the target's signature.
More precisely, the observation model at the $k$th sensor adheres to the following composite hypothesis testing formulation:
\begin{equation}
\begin{cases}
\mathcal{H}_{0}\,:\, & r_{k}=n_{k}\\
\mathcal{H}_{1}\,:\, & r_{k}=\xi_{k}\,g(\bm{p}^{\mathrm{t}},\bm{p}_{k}^{\mathrm{sen}})+n_{k}
\end{cases}\,,\label{eq: local sensor hypothesis testing problem}
\end{equation}
where $r_{k}\in\mathbb{R}$ denotes the received measurement, $n_{k}\sim\mathcal{N}(0,\sigma_{n,k}^{2})$ accounts for local sensing noise, and $\xi_{k}\sim\mathcal{N}(0,\theta)$ captures the spatially uncorrelated, zero-mean Gaussian fluctuations induced by the random nature of the target's emission profile.
The parameter $\theta$, denoting the average power of the emitted signal, is considered known in this work.
Furthermore, the function $g(\bm{p}^{\mathrm{t}},\bm{p}_{k}^{\mathrm{sen}})$ represents a general Amplitude Attenuation Function (AAF), which encodes the deterministic distance- and geometry-dependent path-loss.
The AAF is parameterized by the \emph{unknown} target position $\bm{p}^{\mathrm{t}}\in\mathbb{R}^{d}$ and the known sensor position $\bm{p}_{k}^{\mathrm{sen}}\in\mathbb{R}^{d}$, the latter being obtained via self-localization protocols.

A key structural assumption here is that the sensors are sufficiently spaced such that the additive noise terms $n_{k}$ and the fading coefficients $\xi_{k}$ are \emph{mutually independent} across $k$.
As a result, the marginal pdfs of the sensor observations under both hypotheses can be fully characterized: under $\mathcal{H}_{0}$, we have $r_{k}\,|\,\mathcal{H}_{0}\sim\mathcal{N}(0,\sigma_{n,k}^{2})$, whereas under $\mathcal{H}_{1}$, the sensing model yields $r_{k}\,|\,\mathcal{H}_{1}\sim\mathcal{N}(0,\,\theta\,g^{2}(\bm{p}^{\mathrm{t}},\bm{p}_{k}^{\mathrm{sen}})+\sigma_{n,k}^{2})$.

In this work, we assume that each sensor performs local inference based \emph{solely on its own measurement}, thereby operating under a decoupled decision-making paradigm.
Although the underlying statistical inference problem at each node involves a \emph{composite} hypothesis test (owing to the parametric uncertainty in $\bm{p}^{\mathrm{t}}$), this challenge can be overcome for the adopted signal model.
To this end, we use the Neyman-Pearson criterion~\cite{Kay1998} and formulate the decision rule at the $k$th sensor in terms of the local Log-Likelihood Ratio (LLR), i.e.:
\begin{equation}
\lambda_{k}\triangleq\ln\left[\frac{p(r_{k}\,|\,\mathcal{H}_{1})}{p(r_{k}\,|\,\mathcal{H}_{0})}\right]=\alpha_{k}+\beta_{k}\,r_{k}^{2}\,,
\end{equation}
where the coefficients $\alpha_k$ and $\beta_k$ are defined as
\begin{align}
\alpha_{k}\triangleq & \frac{1}{2}\ln\left[\frac{\sigma_{n,k}^{2}}{\sigma_{n,k}^{2}+\theta\,g^{2}(\bm{p}^{\mathrm{t}},\bm{p}_{k}^{\mathrm{sen}})}\right],\\
\beta_{k}\triangleq & \frac{\theta\,g^{2}(\bm{p}^{\mathrm{t}},\bm{p}_{k}^{\mathrm{sen}})}{\sigma_{n,k}^{2}\left(\sigma_{n,k}^{2}+\theta\,g^{2}(\bm{p}^{\mathrm{t}},\bm{p}_{k}^{\mathrm{sen}})\right)}\,.\nonumber 
\end{align}
Both $\alpha_k$ and $\beta_k$ are strictly positive, implying that $\lambda_k$ is a monotonically increasing function of the received signal energy $r_k^2$, irrespective of the unknown target location $\bm{p}^{\mathrm{t}}$.
Hence, by the Karlin–Rubin theorem \cite[p. 76]{lehmann2022}, the optimal decision rule at each sensor is a simple threshold test: choose $\mathcal{H}_{1}$ if $r_{k}^{2}>\gamma_{k}$, and $\mathcal{H}_0$ otherwise,
where $\gamma_k$ is a tunable threshold.
This test is Uniformly Most Powerful (UMP) in the Neyman–Pearson sense.

Each sensor, therefore, implements a local energy detector~\cite{ciuonzo2017quantizer}, with performance quantified by the detection
\begin{equation}
P_{\mathrm{d},k}\triangleq\Pr\left(r_{k}^{2}\geq\gamma_{k}\,|\,\mathcal{H}_{1}\right)=2\,\mathcal{Q}\left(\sqrt{\frac{\gamma_{k}}{\sigma_{n,k}^{2}+\theta\,g^{2}(\bm{p}^{\mathrm{t}},\bm{p}_{k}^{\mathrm{sen}})}}\right),
\end{equation}
and the false-alarm probabilities
\begin{equation}
P_{\mathrm{f},k}\triangleq\Pr\left(r_{k}^{2}\geq\gamma_{k}\,|\,\mathcal{H}_{0}\right)=2\,\mathcal{Q}\left(\sqrt{\frac{\gamma_{k}}{\sigma_{n,k}^{2}}}\right)\,.
\end{equation}
Following this local inference, each sensor transmits a binary decision $d_{k}\in\left\{ \mathcal{H}_{0},\mathcal{H}_{1}\right\} $ to the FC according to the modulation detailed hereafter.

\noindent
\textbf{Modulation format:} each decision is mapped into $x_{k}\in{\cal X}=\{-1,+1\}$, namely a Binary Phase-Shift Keying (BPSK) modulation: without loss of generality we assume that $d_{k}=\mathcal{H}_{i}$
maps into $x_{k}=(2i-1)$, $i\in\{0,1\}$.

The quality of the WSN is fully characterized by the conditional joint pmfs $\Pr(\bm{x}|\mathcal{H}_{i})$, describing the behavior of the transmitted decision vector $\bm{x}$ under each hypothesis $\mathcal{H}_i$, with $i \in \{0,1\}$. 
Under the target-present hypothesis $\mathcal{H}_1$, the pmf admits the following \emph{factorized} representation (since sensor decisions are conditionally independent):
 \begin{gather}
\Pr(\bm{x}|\mathcal{H}_{1})=\prod_{k=1}^{K}P_{\mathrm{d},k}(\bm{p}^{\mathrm{t}})^{\frac{1+x_{k}}{2}}\cdot[1-P_{\mathrm{d},k}(\bm{p}^{\mathrm{t}})]^{\frac{1-x_{k}}{2}}\,,
\end{gather}
where $P_{\mathrm{d},k}(\bm{p}^{\mathrm{t}})$ denotes the local detection probability of sensor $k$ as a function of the 
unknown target location $\bm{p}^{\mathrm{t}}$. 
An analogous expression holds under the null hypothesis $\mathcal{H}_0$, with the detection probabilities replaced by false alarm probabilities $P_{\mathrm{f},k}$, which are \emph{independent} of the target position. 

We notice that the full specification of the pmf under $\mathcal{H}_1$ is impossible since the target position $\bm{p}^{\mathrm{t}}$ is \emph{unknown}.
This makes the DD problem considered in this work a \emph{composite hypothesis test}~\cite{Kay1998}.

\subsection{Channel Model}\label{subsec:channel_model}

The sensors transmit their local decisions $x_k$'s over a wireless flat-fading multiple access channel~\cite{Ciuonzo2012} to a FC equipped with an RHS. The RHS comprises $M$ reflecting elements and $N$ external receive feeds, each connected to an RF chain. The channel between the $K$ sensors and the RHS is denoted by $\bm{H} \in \mathbb{C}^{M \times K}$, where $\bm{h}_k \in \mathbb{C}^{M}$ is the contribution of the $k$th sensor. The channel between the RHS and the receive feeds is represented by $\bm{G} \in \mathbb{C}^{N \times M}$.

\noindent
\textbf{Sensors-RHS links:}
We adopt a Rician fading model for the link between the $k$th sensor and the RHS:
\begin{align}
\bm{h}_{k} & =\sqrt{P(d_{k}^{\mathrm{sen\rightarrow RHS}},\nu)}\,\left(b_{k}\,\bm{h}_{k}^{\mathrm{LoS}}+\sqrt{1-b_{k}^{2}}\,\hat{\bm{h}}_{k}\right)\,,\label{eq: sensor-RIS channel}
\end{align}
where $P(d,\nu) = \mu (d/d_0)^{-\nu}$ models the path loss as a function of distance $d$, with reference attenuation $\mu$, reference distance $d_0$, and path loss exponent $\nu$. The distance $d_{k}^{\mathrm{sen\rightarrow RHS}}$ is measured from the $k$th sensor to the center of the RHS, assuming far-field conditions.
The line-of-sight component is modeled as:
\begin{equation}
\bm{h}_{k}^{\mathrm{LoS}}=\bm{a}^{\mathrm{upa}}(\theta_{k}^{\mathrm{AoA}},\phi_{k}^{\mathrm{AoA}})\,e^{j\tau_{k}}\,;
\end{equation}
where $\bm{a}^{\mathrm{upa}}(\cdot, \cdot)$ is the steering vector of a uniform planar array:
\begin{equation}
\left[\bm{a}^{\mathrm{upa}}(\theta,\phi)\right]_{m_{x},m_{y}}=\begin{bmatrix}e^{j\frac{2\pi}{\lambda}(m_{x}\,\Delta_{\mathrm{h}}^{\mathrm{rhs}}\sin\theta\cos\phi+m_{y}\,\Delta_{\mathrm{v}}^{\mathrm{rhs}}\sin\theta\sin\phi)}\end{bmatrix},
\end{equation}
with $\lambda$ the wavelength, and $\Delta_{\mathrm{h}}^{\mathrm{rhs}}$ and $\Delta_{\mathrm{v}}^{\mathrm{rhs}}$ denoting the inter-element spacing along the horizontal and vertical axes, respectively. 
The angles $(\theta_{k}^{\mathrm{AoA}}, \phi_{k}^{\mathrm{AoA}})$ capture the azimuth and elevation angles-of-arrival from the $k$th sensor, while $\tau_k \sim \mathcal{U}(0, 2\pi)$ models a random phase offset due to initial delay.
The non-line-of-sight component $\hat{\bm{h}}_k \sim \mathcal{N}_{\mathbb{C}}(\bm{0}_M, \bm{I}_M)$ captures the (normalized) scattered multipath contribution, and $b_k \in [0,1]$ denotes the Rician factor, which quantifies the relative strength of the LoS path.

\noindent
\textbf{RHS-feeds links:} the channel matrix between the RHS and the $N$ feeds follows a \emph{near-field} model, 
based on the well-known (deterministic) \emph{spherical} wave equation form. Specifically, the $(n,m)$th element of $\bm{G}$ equals~\cite{jamali2020intelligent}:
\begin{equation}
g_{n,m}=\left(\frac{\lambda}{4\pi}\right)\sqrt{\eta\,G_{n,m}^{\mathrm{rhs}}G_{n,m}^{\mathrm{fc}}}\,\frac{\exp(-j\,(2\pi/\lambda)\left\Vert \bm{p}_{n}^{\mathrm{fc}}-\bm{p}_{m}^{\mathrm{rhs}}\right\Vert )}{\left\Vert \bm{p}_{n}^{\mathrm{fc}}-\bm{p}_{m}^{\mathrm{rhs}}\right\Vert },
\end{equation}
where $\bm{p}_{n}^{\mathrm{fc}}$ (resp. $\bm{p}_{m}^{\mathrm{rhs}}$) denotes the 3D position of the $n$th feed (resp. $m$th RHS element).
Also, $G_{n,m}^{\mathrm{rhs}}$ denotes the transmit (re-radiation) gain of the $m$th RHS element toward $n$th feed.
Similarly, $G_{n,m}^{\mathrm{fc}}$ characterizes the receive gain at the $n$th antenna for power emitted from the $m$th RHS element.
Finally, $\eta$ underlines the efficiency of the RHS.
The (RHS) re-radiation gain $G_{n,m}^{\mathrm{rhs}}$ has the following form:
\begin{align}
\ensuremath{G_{n,m}^{\mathrm{rhs}}} & =(4\pi\,/\lambda^{2})\,A_{n,m}^{\mathrm{rhs}}=(4\pi\,/\lambda^{2})\cdot\left(\Delta_{\mathrm{h}}^{\mathrm{rhs}}\Delta_{\mathrm{v}}^{\mathrm{rhs}}\right)\cdot\rho_{n,m}^{\mathrm{rhs}}\,,
\end{align}
where $A_{n,m}^{\mathrm{rhs}}$ denotes the aperture of $m$th RHS element as seen by the $n$th feed.  
Also, $\Delta_{\mathrm{h}}^{\mathrm{rhs}}$ (resp. $\Delta_{\mathrm{v}}^{\mathrm{rhs}}$)
denotes the RHS element size along the horizontal (resp. vertical) axis, while $\rho_{n,m}^{\mathrm{rhs}}$ denotes the corresponding directivity factor.
Similarly, the (feed) receive gain is:
\begin{align}
\ensuremath{G_{n,m}^{\mathrm{fc}}} & =(4\pi\,/\lambda^{2})\,A_{n,m}^{\mathrm{fc}}=(4\pi\,/\lambda^{2})\cdot\left(\Delta_{\mathrm{h}}^{\mathrm{fc}}\Delta_{\mathrm{v}}^{\mathrm{fc}}\right)\cdot\rho_{n,m}^{\mathrm{fc}}\,,
\end{align}
where $A_{n,m}^{\mathrm{fc}}$ denotes the aperture of $m$th RHS element as seen by the $n$th feed. 
Also, $\Delta_{\mathrm{h}}^{\mathrm{fc}}$ (resp. $\Delta_{\mathrm{v}}^{\mathrm{fc}}$)
denotes the FC array element size along the horizontal (resp. vertical) axis, while $\rho_{n,m}^{\mathrm{fc}}$ denotes the corresponding directivity factor.
In both the gain terms, the directivity factors $\rho_{n,m}^{\mathrm{rhs}}$ and $\rho_{n,m}^{\mathrm{fc}}$ are obtained via the functional form $\rho(\theta,\phi)$, which has the following expression~\cite{jamali2020intelligent,ellingson2021path}:
\begin{align}
\rho(\theta,\phi) & \triangleq\begin{cases}
2\,(2q+1)\cos^{2q}(\theta) & \begin{array}{c}
0\leq\theta\leq\pi/2\,\\
0\leq\phi\leq2\pi
\end{array},\\
0 & \mathrm{otherwise}
\end{cases}\label{eq: general directivity pattern}
\end{align}
where $q\geq 0 $ is a real number that determines the directivity, while the multiplication factor $2\,(2q+1)$ ensures energy conservation ($\frac{1}{4\pi}\int_{\Omega}\rho(\theta,\phi)\,d\Omega=1$).
The same term also represents the value achieved in the direction of maximum gain~\cite{tang2020wireless}.
The aforementioned pattern expressions can be calculated as~\cite{abrardo2021intelligent,DegliEsposti2022,feng2023near}:
\begin{align}
\rho_{n,m}^{\mathrm{rhs}} & =2\,(2q+1)\,\cos^{2q}(\theta_{n,m}^{\mathrm{rhs}})\,;\\
\rho_{n,m}^{\mathrm{fc}} & =2\,(2q+1)\,\cos^{2q}(\theta_{n,m}^{\mathrm{fc}})\,,
\end{align}
where 
\begin{align}
\cos(\theta_{n,m}^{\mathrm{rhs}})= & \frac{(\bm{p}_{n}^{\mathrm{fc}}-\bm{p}_{m}^{\mathrm{rhs}})^{T}\,\bm{u}^{\mathrm{rhs}}}{\left\Vert \bm{p}_{n}^{\mathrm{fc}}-\bm{p}_{m}^{\mathrm{rhs}}\right\Vert }\,,\\
\cos(\theta_{n,m}^{\mathrm{fc}})= & \frac{(\bar{\bm{p}}^{\mathrm{rhs}}-\bm{p}_{n}^{\mathrm{fc}})^{T}\,\left(\bm{p}_{m}^{\mathrm{rhs}}-\bm{p}_{n}^{\mathrm{fc}}\right)}{\left\Vert \bar{\bm{p}}^{\mathrm{rhs}}-\bm{p}_{n}^{\mathrm{fc}}\right\Vert \left\Vert \bar{\bm{p}}^{\mathrm{rhs}}-\bm{p}_{n}^{\mathrm{fc}}\right\Vert }\,.
\end{align}
In other terms, $\cos(\theta_{n,m}^{\mathrm{rhs}})$ represents the cosine angle between the RHS element boresight direction and the vector denoting the reflection direction of the $m$th RHS element toward $n$th receive feed.  
Conversely, $\cos(\theta_{n,m}^{\mathrm{fc}})$ represents the cosine angle between the same reflection direction (but in the coordinate system of $n$th feed) and the vector of maximum receive gain. 
In this work, we assume that $N$ receive antennas are aligned with their maximum gain directions towards the geometrical center of the RHS ($\bar{\bm{p}}^{\mathrm{rhs}}$), namely a full-illumination configuration~\cite{jamali2020intelligent}.

The received signal vector at the FC, denoted by $\bm{y}\in\mathbb{C}^{N}$, admits the compact expression:
\begin{align}
\bm{y}= & \left(\bm{G}\bm{\Theta}\bm{H}\right)\bm{D}_{\alpha}\,\bm{x}+\bm{w}=\bm{H}^{e}(\bm{\Theta})\,\bm{D}_{\alpha}\,\bm{x}+\bm{w},\label{eq: signal_model}
\end{align}
where $\bm{x}\in\mathcal{X}^{K}$ is the transmitted symbol vector, and $\bm{w}\sim\mathcal{N}_{\mathbb{C}}(\bm{0}_{N},\ensuremath{\sigma_{w}^{2}\bm{I}_{N})}$ models additive complex Gaussian noise.
The diagonal matrix $\bm{\Theta}=\mathrm{diag}(e^{j\varphi_{1}},\ldots,e^{j\varphi_{M}})$ encodes the tunable phase shifts of the $M$-element RHS, while $\bm{D}_{\alpha}=\mathrm{diag}(\alpha_{1},\ldots,\alpha_{K})$ captures the transmit energy at the $K$ sensor nodes, with $\alpha_{k}\in\mathbb{R}^{+},\,\forall k \in \mathcal{K}$.
The effective channel matrix is defined as $\bm{H}^{e}(\bm{\Theta})\,\triangleq\left(\bm{G}\bm{\Theta}\bm{H}\right)$ representing the end-to-end cascaded channel influenced by the RHS configuration.

\subsection{Statistical Characterization}\label{subsec:stat_char}

Under the statistical assumptions introduced in Secs.~\ref{subsec:sensing_model} and~\ref{subsec:channel_model}, the received signal vector $\bm{y}$, conditioned on hypothesis $\mathcal{H}_i$, admits the following second-order statistical characterization.
Spefifically, due to the potential impropriety of $\bm{y}|\mathcal{H}_i$, this characterization must include \emph{both} the covariance and the pseudo-covariance matrices:
\begin{align}
\ensuremath{\mathbb{E}\{\bm{y}|\mathcal{H}_{i}\}} & =\bm{H}^{e}(\bm{\Theta})\,\bm{D}_{\alpha}\,(2\,\bm{\rho}_{i}-\bm{1}_{K})\,;\label{eq:2nd_order_char}\\
\ensuremath{\mathrm{Cov}(\bm{y}|\mathcal{H}_{i})} & \ensuremath{=}\bm{H}^{e}(\bm{\Theta})\,\bm{D}_{\alpha}\,\mathrm{Cov}\left(\bm{x}|\mathcal{H}_{i}\right)\,\bm{D}_{\alpha}\,\bm{H}^{e}(\bm{\Theta})^{\dagger}+\sigma_{w}^{2}\,\bm{I}_{N}\,;\nonumber \\
\mathrm{PCov}(\bm{y}|\mathcal{H}_{i}) & =\bm{H}^{e}(\bm{\Theta})\,\bm{D}_{\alpha}\,\mathrm{Cov}(\bm{x}|\mathcal{H}_{i})\,\bm{D}_{\alpha}\,\bm{H}^{e}(\bm{\Theta})^{T}\,.\nonumber 
\end{align}
In the above, the vectors $\ensuremath{\bm{\rho}_{1}(\bm{p}^{\mathrm{t}})\triangleq[P_{\mathrm{d},1}(\bm{p}^{\mathrm{t}})\cdots P_{\mathrm{d},K}(\bm{p}^{\mathrm{t}})]^{T}}$ and $\bm{\rho}_{0}\triangleq[P_{\mathrm{f},1}\cdots P_{\mathrm{f},K}]^{T}$ respectively collect the detection and false-alarm probabilities associated with the $K$ sensors. 
The matrix $\mathrm{Cov}\left(\bm{x}|\mathcal{H}_{i}\right)$ denotes the hypothesis-conditioned covariance of the binary decision vector $\bm{x}$.
Under the sensing model introduced in Sec.~\ref{subsec:sensing_model}, this covariance takes the following form under $\mathcal{H}_{1}$:
\begin{equation}
\mathrm{Cov}\left(\bm{x}|\mathcal{H}_{1}\right)=\underbrace{4\,\mathrm{diag}\left(\begin{array}{c}
P_{\mathrm{d},1}(\bm{p}^{\mathrm{t}})[1-P_{\mathrm{d},1}(\bm{p}^{\mathrm{t}})]\\
\vdots\\
P_{\mathrm{d},K}(\bm{p}^{\mathrm{t}})[1-P_{\mathrm{d},K}(\bm{p}^{\mathrm{t}})]
\end{array}\right)}_{\triangleq\,\bm{C}_{1}(\bm{\rho}_{1}(\bm{p}^{t}))}\,,
\end{equation}
where the last definition highlights that $\mathrm{Cov}\left(\bm{x}|\mathcal{H}_{1}\right)$ depends on the target position \emph{only} via the detection probability vector $\bm{\rho}_1(\bm{p}^{\mathrm{t}})$.
An analogous expression holds for $\mathrm{Cov}\left(\bm{x}|\mathcal{H}_{0}\right)$, obtained by replacing each $P_{\mathrm{d},k}(\bm{p}^{\mathrm{t}})$ with the corresponding $P_{\mathrm{f},k}$.

As a direct implication, $\ensuremath{\mathrm{Cov}(\bm{y}|\mathcal{H}_{1})}$ and $\ensuremath{\mathrm{PCov}(\bm{y}|\mathcal{H}_{1})}$ are inherently \emph{dependent} on the (unknown) target position $\bm{p}^{\mathrm{t}}$, whereas $\ensuremath{\mathrm{Cov}(\bm{y}|\mathcal{H}_{0})}$ and $\ensuremath{\mathrm{PCov}(\bm{y}|\mathcal{H}_{0})}$ are \emph{invariant} with respect to it. This asymmetry in the statistical structure of the received signal across hypotheses will later play a key role in the design of fusion strategies.

The WSN system performance is evaluated in terms of the global probabilities of \textit{false alarm} $P_{F_{0}}\triangleq\Pr(\Lambda>\gamma|H_{0})$ and  \textit{detection} $P_{D_{0}}\triangleq\Pr(\Lambda>\gamma|H_{1})$.

\section{Joint Fusion Rule and RHS design Problem Formulation}\label{sec_fusion_rule_RIS}

The goal of this work is to develop a practical design framework that enables high-performance, goal-oriented detection of a target located at unknown position $\bm{p}^\mathrm{t}$, via the joint optimization of the fusion rule statistic $\Lambda = \mathcal{F}(\bm{y})$ and the RHS matrix $\bm{\Theta}$. 
To this end, we first revisit the GLR fusion rule under a fixed RHS setup and highlight the limitations that preclude its practical implementation (Sec.~\ref{subsec: optimal_LLR}).
We then introduce two tractable design approaches--eFuC and bFuC--based on WL processing of the received signal vector (Sec.~\ref{subsec: WLlinear_design}).

\subsection{Optimal fusion rule}\label{subsec: optimal_LLR}
The optimal detector is the (global) LLR test:
\begin{equation}
\left\{ \Lambda_{\mathrm{OPT}}\triangleq\ln\left[\frac{p(\bm{y}|\mathcal{H}_{1})}{p(\bm{y}|\mathcal{H}_{0})}\right]\right\} \begin{array}{c}
{\scriptstyle \hat{\mathcal{H}}=\mathcal{H}_{1}}\\
\gtrless\\
{\scriptstyle \hat{\mathcal{H}}=\mathcal{H}_{0}}
\end{array}\gamma_{\mathrm{fc}}\,.\label{eq:neyman_pearson_test}
\end{equation}
Here, $\Lambda_{\mathrm{OPT}}$ denotes the LLR, $\gamma_{\mathrm{fc}}$ is the decision threshold, and $\hat{\mathcal{H}}$ is the selected hypothesis.
The threshold $\gamma$ can be set to control the false-alarm rate or minimize fusion error probability~\cite{Kay1998}.
Unfortunately, the LLR \emph{cannot be implemented} because of unavailability of target position $\bm{p}^{\mathrm{t}}$.
Hence, in this case, a GLR statistic~\cite{Kay1998} seems more appropriate
\begin{gather}
\Lambda_{\mathrm{GLR}}=\ln\left[\frac{\max_{\bm{p}^{\mathrm{t}}\in\mathcal{A}}\sum_{\bm{x}\in{\cal X}^{K}}p(\bm{y}|\bm{x})f_{1}(\bm{x};\bm{p}^{\mathrm{t}})}{\sum_{\bm{x}\in{\cal X}^{K}}p(\bm{y}|\bm{x})\,f_{0}(\bm{x})}\right]\label{eq:GLR_RIS}\\
=\max_{\bm{p}^{\mathrm{t}}\in\mathcal{A}}\,\ln\left[\frac{\sum_{\bm{x}\in{\cal X}^{K}}\exp\left(-\frac{\bm{\|y}-\bm{H}^{e}(\bm{\Theta})\bm{D}_{\alpha}\bm{x}\|^{2}}{\sigma_{w}^{2}}\right)f_{1}(\bm{x};\bm{p}^{\mathrm{t}})}{\sum_{\bm{x}\in{\cal X}^{K}}\exp\left(-\frac{\bm{\|y}-\bm{H}^{e}(\bm{\Theta})\bm{D}_{\alpha}\bm{x}\|^{2}}{\sigma_{w}^{2}}\right)f_{0}(\bm{x})}\right]\,.\nonumber 
\end{gather}
For notational compactness, in Eq.~\eqref{eq:GLR_RIS} we have denoted the resulting likelihood functions as $f_{1}(\bm{x};\bm{p}^{\mathrm{t}})$ and $f_{0}(\bm{x})$, corresponding to hypotheses $\mathcal{H}_1$ and $\mathcal{H}_0$, respectively.
Although the GLR accounts for the unknown target location,
it ($i$) incurs \emph{exponential complexity} ($\propto2^{K}$), ($ii$) requires a search over $\bm{p}^{\mathrm{t}}$, which is typically implemented by discretizing the surveillance region $\mathcal{A}$ into $N_{\mathrm{t}}$ candidate points, and ($iii$) lacks a closed-form performance characterization, thus hindering its use for RHS design purposes.

 \subsection{Deflection-based Design of RHS Matrix \& WL Fusion Rule}\label{subsec: WLlinear_design}
 Similarly to~\cite{mudkey2022wireless,Ciuonzo2025IoT}, in this work we adopt \emph{deflection measures} as the relevant design metric:
\begin{gather}
D_{i}\left(\Lambda\right)\triangleq\left(\mathbb{E}\{\Lambda|\mathcal{H}_{1}\}-\mathbb{E}\{\Lambda|\mathcal{H}_{0}\}\right)^{2}\,/\,\mathrm{var}\{\Lambda|\mathcal{H}_{i}\}\,,
\end{gather}
where $D_{0}(\cdot)$ and $D_{1}(\cdot)$
correspond to the \emph{normal}~\cite{Picinbono1995} and \emph{modified}~\cite{Quan2008}
deflections, respectively.
Indeed, deflection metrics represent a flexible and manageable design tool when the fusion rule is constrained to be a WL fusion statistic, i.e.,
\begin{equation}
\Lambda_{\mathrm{WL}}=\underline{\bm{a}}^{\dagger}\bm{\underline{y}}+b\,.\label{eq: WL fusion statistic}
\end{equation}
The statistic $\Lambda_{\mathrm{WL}}$ is then used to implement the hypothesis test similarly to as Eq.~\eqref{eq:neyman_pearson_test}.
In that case, the expression of the deflection measures simplify as:
\begin{gather}
D_{i}(\Lambda_{\mathrm{WL}})\triangleq\frac{\left(\underbar{\ensuremath{\bm{a}}}^{\dagger}\left(\mathbb{E}\{\text{\ensuremath{\underbar{\ensuremath{\bm{y}}}}}|\mathcal{H}_{1}\}-\mathbb{E}\{\text{\ensuremath{\underbar{\ensuremath{\bm{y}}}}}|\mathcal{H}_{0}\}\right)\right)^{2}}{\underbar{\ensuremath{\bm{a}}}^{\dagger}\mathrm{Cov}(\text{\ensuremath{\underbar{\ensuremath{\bm{y}}}}}|\mathcal{H}_{i})\,\underbar{\ensuremath{\bm{a}}}}\,.
\end{gather}
The use of WL fusion is motivated by its reduced complexity and the improper nature of $\bm{y}|\mathcal{H}_{i}$, characterized by $\mathrm{PCov}(\bm{y}|\mathcal{H}_{i})\neq\bm{O}_{N\times N}$  (see third line of Eq.~\eqref{eq:2nd_order_char}), and is further supported by its competitive performance in analogous WSN scenarios~\cite{ciuonzo2015,ciuonzo2017rician}.
 
In what follows, we consider \emph{design strategies} based on the fully-complete second order characterization (FuC) reported in Eq.~\eqref{eq:2nd_order_char}.
By making the appropriate substitutions, the deflection measures can be rewritten as:
\begin{align}
D_{\mathrm{\mathrm{FuC},}1}\left(\text{\ensuremath{\underbar{\ensuremath{\bm{a}}}}},\bm{\Theta};\bm{p}^{\mathrm{t}}\right) & =\frac{4\cdot\left(\underbar{\ensuremath{\bm{a}}}^{\dagger}\left[\underline{\bm{H}}^{e}(\bm{\Theta})\,\bm{D}_{\alpha}\,\ensuremath{\bm{\rho}_{10}}(\bm{p}^{\mathrm{t}})\right]\right)^{2}}{\underbar{\ensuremath{\bm{a}}}^{\dagger}\mathrm{Cov}(\text{\ensuremath{\underbar{\ensuremath{\bm{y}}}}}|\mathcal{H}_{1};\bm{p}^{\mathrm{t}})\,\underbar{\ensuremath{\bm{a}}}};\label{eq: deflection_WL}\\
D_{\mathrm{\mathrm{FuC},}0}\left(\text{\ensuremath{\underbar{\ensuremath{\bm{a}}}}},\bm{\Theta};\bm{p}^{\mathrm{t}}\right) & =\frac{4\cdot\left(\underbar{\ensuremath{\bm{a}}}^{\dagger}\left[\underline{\bm{H}}^{e}(\bm{\Theta})\,\bm{D}_{\alpha}\,\ensuremath{\bm{\rho}_{10}}(\bm{p}^{\mathrm{t}})\right]\right)^{2}}{\underbar{\ensuremath{\bm{a}}}^{\dagger}\mathrm{Cov}(\text{\ensuremath{\underbar{\ensuremath{\bm{y}}}}}|\mathcal{H}_{0})\,\underbar{\ensuremath{\bm{a}}}};
\end{align}
where $\bm{\rho}_{10}(\bm{p}^{\mathrm{t}})\triangleq\,(\bm{\rho}_{1}(\bm{p}^{\mathrm{t}})-\bm{\rho}_{0})$ while the \emph{augmented covariance} has the following expression:
\begin{equation}
\mathrm{Cov}(\text{\ensuremath{\underbar{\ensuremath{\bm{y}}}}}|\mathcal{H}_{i})=\underline{\bm{H}}^{e}(\bm{\Theta})\,\bm{D}_{\alpha}\,\mathrm{Cov}(\bm{x}|\mathcal{H}_{i})\,\bm{D}_{\alpha}\,\underline{\bm{H}}^{e}(\bm{\Theta})^{\dagger}+\sigma_{w}^{2}\,\bm{I}_{2N}\,.\label{eq: augmented covariance rx signal vector}
\end{equation}
Unfortunately, despite simplifications originating from WL assumption in fusion rule design, both the deflection measures in Eq.~\eqref{eq: deflection_WL} \emph{depend} on $\bm{p}^{\mathrm{t}}$ and thus they can lead to \emph{non-realizable} design solutions.
To cope with this problem, we pursue \emph{two different options}.

In the \emph{first case}, referred to as eFuC, since the deflections depend on the target position only via the detection probability vector $\bm{\rho}_{1}(\bm{p}^{\mathrm{t}})$, we suggest to replace this vector with its \emph{expectation}, namely:
\begin{equation}
\ensuremath{\bar{\bm{\rho}}_{1}\triangleq\begin{bmatrix}\int_{\mathcal{A}}P_{\mathrm{d},1}(\bm{p}^{\mathrm{t}})\,f(\bm{p}^{\mathrm{t}})\,d\bm{p}^{\mathrm{t}}\\
\vdots\\
\int_{\mathcal{A}}P_{\mathrm{d},K}(\bm{p}^{\mathrm{t}})\,f(\bm{p}^{\mathrm{t}})\,d\bm{p}^{\mathrm{t}}
\end{bmatrix}},
\end{equation}
where $\,f(\bm{p}^{\mathrm{t}})$ represents the location pdf of the target.
If this is unknown or hard to estimate, a simply uniform distribution
over the surveillance area $\mathcal{A}$ can be used.
Upon replacing $\bm{\rho}_{1}(\bm{p}^{\mathrm{t}})$ with $\bar{\bm{\rho}}_{1}$, the relevant deflection measures to be optimized become:
\begin{align}
D_{\mathrm{\mathrm{eFuC},}1}\left(\text{\ensuremath{\underbar{\ensuremath{\bm{a}}}}},\bm{\Theta}\right) & =\frac{4\cdot\left(\underbar{\ensuremath{\bm{a}}}^{\dagger}\left[\underline{\bm{H}}^{e}(\bm{\Theta})\,\bm{D}_{\alpha}\,\ensuremath{\bar{\bm{\rho}}_{10}}\right]\right)^{2}}{\underbar{\ensuremath{\bm{a}}}^{\dagger}\overline{\mathrm{Cov}}(\text{\ensuremath{\underbar{\ensuremath{\bm{y}}}}}|\mathcal{H}_{1})\,\underbar{\ensuremath{\bm{a}}}}\,;\label{eq: deflection_mFuC1}\\
D_{\mathrm{\mathrm{eFuC},}0}\left(\text{\ensuremath{\underbar{\ensuremath{\bm{a}}}}},\bm{\Theta}\right) & =\frac{4\cdot\left(\underbar{\ensuremath{\bm{a}}}^{\dagger}\left[\underline{\bm{H}}^{e}(\bm{\Theta})\,\bm{D}_{\alpha}\,\ensuremath{\bar{\bm{\rho}}_{10}}\right]\right)^{2}}{\underbar{\ensuremath{\bm{a}}}^{\dagger}\mathrm{Cov}(\text{\ensuremath{\underbar{\ensuremath{\bm{y}}}}}|\mathcal{H}_{0})\,\underbar{\ensuremath{\bm{a}}}}\,;\label{eq: deflection_mFuC0}
\end{align}
where $\bar{\bm{\rho}}_{10}\triangleq\,(\bar{\bm{\rho}}_{1}-\bm{\rho}_{0})$ and
\begin{equation}
\overline{\mathrm{Cov}}(\text{\ensuremath{\underbar{\ensuremath{\bm{y}}}}}|\mathcal{H}_{1})=\underline{\bm{H}}^{e}(\bm{\Theta})\,\bm{D}_{\alpha}\bm{C}_{1}(\bar{\bm{\rho}}_{1})\bm{D}_{\alpha}\,\underline{\bm{H}}^{e}(\bm{\Theta})^{\dagger}+\sigma_{w}^{2}\bm{I}_{2N}\,.\label{eq: augmented_covariance_mFuC}
\end{equation}
In the \emph{second case}, referred to as bFuC, to cope with uncertainty in the target position, we propose the adoption of a WL filter-bank:
\begin{equation}
\Lambda_{\mathrm{bWL}}=\max_{j=1,\cdots N_{\mathrm{t}}}\underline{\bm{a}}_{j}^{\dagger}\bm{\underline{y}}+b_{j}\,.\label{eq: FB-WL fusion statistic}
\end{equation}
By doing so, each different filter $\underline{\bm{a}}_{j}$ can be chosen to optimize the deflection for a specific candidate target position $\bm{p}^{\mathrm{t}}[j]$, namely $D_{\mathrm{\mathrm{FuC},}i}\left(\text{\ensuremath{\underbar{\ensuremath{\bm{a}}}_{j}}},\bm{\Theta};\bm{p}^{\mathrm{t}}=\bm{p}^{\mathrm{t}}[j]\right)$. 
Still, the optimization is coupled because of the common dependence on the phase shift matrix $\bm{\Theta}$. Hence, in such a case, we propose to maximize the
\emph{average} deflection response over all the probed positions:
\begin{gather}
D_{\mathrm{\mathrm{bFuC},}1}\left(\left\{ \text{\ensuremath{\underbar{\ensuremath{\bm{a}}}}}_{j}\right\} _{j=1}^{N_{\mathrm{t}}},\bm{\Theta}\right)=\label{eq: deflection_bFuC1}\\
\sum_{j=1}^{N_{\mathrm{t}}}\frac{4\cdot\left(\underbar{\ensuremath{\bm{a}}}_{j}^{\dagger}\left[\underline{\bm{H}}^{e}(\bm{\Theta})\,\bm{D}_{\alpha}\,\ensuremath{\ensuremath{\bm{\rho}_{10}}(\bm{p}^{\mathrm{t}}[j])}\right]\right)^{2}}{N_{\mathrm{t}}\,\underbar{\ensuremath{\bm{a}}}_{j}^{\dagger}\mathrm{Cov}(\text{\ensuremath{\underbar{\ensuremath{\bm{y}}}}}|\mathcal{H}_{1};\bm{p}^{\mathrm{t}}[j])\,\underbar{\ensuremath{\bm{a}}}_{j}}\,;\nonumber \\
D_{\mathrm{\mathrm{bFuC},}0}\left(\left\{ \text{\ensuremath{\underbar{\ensuremath{\bm{a}}}}}_{j}\right\} _{j=1}^{N_{\mathrm{t}}},\bm{\Theta}\right)=\nonumber \\
\sum_{j=1}^{N_{\mathrm{t}}}\frac{4\cdot\left(\underbar{\ensuremath{\bm{a}}}_{j}^{\dagger}\left[\underline{\bm{H}}^{e}(\bm{\Theta})\,\bm{D}_{\alpha}\,\ensuremath{\ensuremath{\bm{\rho}_{10}}(\bm{p}^{\mathrm{t}}[j])}\right]\right)^{2}}{N_{\mathrm{t}}\,\underbar{\ensuremath{\bm{a}}}_{j}^{\dagger}\mathrm{Cov}(\text{\ensuremath{\underbar{\ensuremath{\bm{y}}}}}|\mathcal{H}_{0})\,\underbar{\ensuremath{\bm{a}}}_{j}}\,.\label{eq: deflection_bFuC0}
\end{gather}
If prior information on a non-uniform target‐location distribution $f(\bm{p}^{\mathrm{t}})$ is available, the objectives in Eqs.~\eqref{eq: deflection_bFuC1} and \eqref{eq: deflection_bFuC1} can be reformulated as a \emph{weighted sum}, where each candidate position is weighted by its corresponding prior probability.

Accordingly, the goal of this work is \emph{twofold}: ($i$) jointly optimize the WL vector $\underbar{\ensuremath{\bm{a}}}$ and phase-shift matrix $\bm{\Theta}$ to maximize the deflections of the eFuC case (namely Eqs.~\eqref{eq: deflection_mFuC1}-\eqref{eq: deflection_mFuC0}); ($ii$) jointly optimize the filter-bank vectors $\left\{ \text{\ensuremath{\underbar{\ensuremath{\bm{a}}}}}_{j}\right\} _{j=1}^{N_{\mathrm{t}}}$ and $\bm{\Theta}$ to maximize the deflections of the bFuC case (namely Eqs.~\eqref{eq: deflection_bFuC1}-\eqref{eq: deflection_bFuC0}).
The resulting designs obtained from Eqs.~\eqref{eq: deflection_mFuC1} and \eqref{eq: deflection_mFuC0} are referred to as \textbf{eFuC-1} and \textbf{eFuC-0}, respectively, while those from Eqs.~\eqref{eq: deflection_bFuC1} and \eqref{eq: deflection_bFuC0} are denoted as \textbf{bFuC-1} and \textbf{bFuC-0}, respectively.

Hence, the resulting optimization problem for eFuC is formulated as:
\vspace{6pt}
\noindent
\begin{tcolorbox}[
    colback=lightbluebox!45,
    colframe=blue,
    boxrule=1.4pt,
    sharp corners,
    left=0pt,         
    right=6pt,
    top=0pt,          
    bottom=4pt,
    boxsep=0pt,
    enhanced,
    before skip=0pt,
    after skip=0pt,
    valign=center,
]

{\hspace*{-0.4pt}
 \raisebox{-\depth}{\color{white}\setlength{\fboxsep}{1pt}%
 \colorbox{blue}{\textbf{Joint RHS \& Fusion Rule design via eFuC criterion}}}}

\vspace{6pt}
\noindent
\begin{equation}
\mathcal{P}_{\mathrm{\mathrm{eFuC}},i}\,:\begin{array}{c}
\underset{\underbar{\ensuremath{\bm{a}}}\,,\bm{\Theta}}{\mathrm{maximize}}\;\frac{\left(\underbar{\ensuremath{\bm{a}}}^{\dagger}\left[\underline{\bm{H}}^{e}(\bm{\Theta})\,\bm{D}_{\alpha}\,\ensuremath{\bar{\bm{\rho}}_{10}}\right]\right)^{2}}{\underbar{\ensuremath{\bm{a}}}^{\dagger}\overline{\mathrm{Cov}}(\text{\ensuremath{\underbar{\ensuremath{\bm{y}}}}}|\mathcal{H}_{i})\,\underbar{\ensuremath{\bm{a}}}}\\
\mathrm{subject\,to}\begin{array}{c}
\left\Vert \underbar{\ensuremath{\bm{a}}}\right\Vert =1\\
\bm{\Theta}=\mathrm{diag}(e^{j\varphi_{1}},\ldots,e^{j\varphi_{M}})
\end{array}
\end{array}
\end{equation}

\end{tcolorbox}
\vspace{4pt}
Differently, for bFuC:
\vspace{4pt}
\noindent
\begin{tcolorbox}[
    colback=lightpurplebox!45,
    colframe=titlepurple,
    boxrule=1.4pt,
    sharp corners,
    left=0pt,         
    right=6pt,
    top=0pt,          
    bottom=4pt,
    boxsep=0pt,
    enhanced,
    before skip=0pt,
    after skip=0pt,
    valign=center,
]

{\hspace*{-0.4pt}
 \raisebox{-\depth}{\color{white}\setlength{\fboxsep}{1pt}%
 \colorbox{titlepurple}{\textbf{Joint RHS \& Fusion Rule design via bFuC criterion}}}}

\vspace{4pt}

\begin{equation}
\mathcal{P}_{\mathrm{\mathrm{bFuC}},i}:\begin{array}{c}
\underset{\left\{ \underbar{\ensuremath{\bm{a}}}_{j}\right\} _{j=1}^{N_{\mathrm{t}}}\,,\bm{\Theta}}{\mathrm{maximize}}\;\sum_{j=1}^{N_{\mathrm{t}}}\frac{\left(\underbar{\ensuremath{\bm{a}}}_{j}^{\dagger}\left[\underline{\bm{H}}^{e}(\bm{\Theta})\,\bm{D}_{\alpha}\,\ensuremath{\bm{\rho}_{10}}(\bm{p}^{\mathrm{t}}[j])\right]\right)^{2}}{\underbar{\ensuremath{\bm{a}}}_{j}^{\dagger}\,\mathrm{Cov}(\text{\ensuremath{\underbar{\ensuremath{\bm{y}}}}}|\mathcal{H}_{i};\bm{p}^{\mathrm{t}}[j])\,\underbar{\ensuremath{\bm{a}}}_{j}}\\
\mathrm{subject\,to}\begin{array}{c}
\left\{ \left\Vert \underbar{\ensuremath{\bm{a}}}_{j}\right\Vert \right\} _{j=1}^{N_{\mathrm{t}}}=1\\
\bm{\Theta}=\mathrm{diag}(e^{j\varphi_{1}},\ldots,e^{j\varphi_{M}})
\end{array}
\end{array}
\end{equation}

\end{tcolorbox}
\vspace{4pt}
Differently from conventional channel-aware DD with standard (viz. fully-digital) arrays, the unit-modulus constraint on $\bm{\Theta}$--introduced by the holographic DF system--induces non-convexity that, coupled with the non-convex objective, renders both  $\mathcal{P}_{\mathrm{\mathrm{eFuC}},i}$ and $\mathcal{P}_{\mathrm{\mathrm{bFuC}},i}$ problems inherently \emph{non-convex}.

\noindent
\textbf{Insight on bias selection:} The deflection metrics associated with eFuC design (Eqs.~\eqref{eq: deflection_mFuC1} and \eqref{eq: deflection_mFuC0}) are independent of the bias term $b$ in the WL statistic~\eqref{eq: WL fusion statistic}.
The same reasoning applies for bFuC deflection measures (Eqs.~\eqref{eq: deflection_bFuC1}~and~\eqref{eq: deflection_bFuC0}) when referring to the bias terms $\left\{ b_{j}\right\} _{j=1}^{N_{\mathrm{t}}}$ in the filter bank WL statistic \eqref{eq: FB-WL fusion statistic}. 
While the scalar bias can be absorbed in the decision threshold $\gamma_{\mathrm{fc}}$, in the case of filter-bank processing different bias terms \emph{can influence the collective behaviour of the bank}.
Under ($a$) Gaussian improperness and ($b$) common augmented covariance assumptions, the GLR leads to a structure matching Eq.~\eqref{eq: FB-WL fusion statistic}, with the corresponding bias terms derived as
\begin{gather}
b_{ij}=-\frac{1}{2}\mathbb{E}\{\underline{\bm{y}}|\mathcal{H}_{1};\bm{p}^{\mathrm{t}}[j]\}^{\dagger}\mathrm{Cov}^{-1}(\text{\ensuremath{\underbar{\ensuremath{\bm{y}}}}}|\mathcal{H}_{i})\,\mathbb{E}\{\underline{\bm{y}}|\mathcal{H}_{1};\bm{p}^{\mathrm{t}}[j]\}\nonumber \\
+\frac{1}{2}\mathbb{E}\{\underline{\bm{y}}|\mathcal{H}_{0}\}^{\dagger}\mathrm{Cov}^{-1}(\text{\ensuremath{\underbar{\ensuremath{\bm{y}}}}}|\mathcal{H}_{i})\,\mathbb{E}\{\underline{\bm{y}}|\mathcal{H}_{0}\}\,.\label{eq: bias_terms_GLR-FB}
\end{gather}
Therefore, after optimizing $\left\{ \text{\ensuremath{\underbar{\ensuremath{\bm{a}}}}}_{j}\right\} _{j=1}^{N_{\mathrm{t}}}$ and $\bm{\Theta}$, we propose to set the biases according to Eq.~\eqref{eq: bias_terms_GLR-FB}.

 
\section{Proposed AO-based Solution}\label{sec:AO_based_solution}

In this work, we adopt the AO approach to efficiently tackle the problems $\mathcal{P}_{\mathrm{eFuC},i}$ and $\mathcal{P}_{\mathrm{bFuC},i}$.
The core idea is to decompose the original problem into two coupled subproblems, which are iteratively optimized in an alternating fashion: one with respect to the fusion weight vector $\underline{\bm{a}}$ (or the set $\{\underline{\bm{a}}_j\}_{j=1}^{N_{\mathrm{t}}}$ in the bFuC case), and the other with respect to the RHS matrix $\bm{\Theta}$.
These two steps are detailed in Sec.~\ref{subsec: fusion_rule+RHS_AO_design}.

It is worth noting that AO is a particular instance of block coordinate descent, known for its broad applicability and strong empirical performance, especially in handling nonconvex problems. Although it does not guarantee global optimality, it often yields high-quality suboptimal solutions in practice. 
In the remainder of this section, we outline the AO-based procedure developed for both eFuC and bFuC formulations.
The section ends with complexity analysis of the design (Sec.~\ref{subsec:complexity}).

\subsection{AO Designed Steps} \label{subsec: fusion_rule+RHS_AO_design}
\noindent
\textbf{Step (A) - Fusion rule design:} 
We first focus on the optimization of the WL vector $\underbar{\ensuremath{\bm{a}}}$ (or the set of WL vectors $\{\underline{\bm{a}}_j\}_{j=1}^{N_{\mathrm{t}}}$) for a \emph{fixed} matrix $\bm{\Theta}_{\mathrm{fix}}$. 
Accordingly, the WL vector design problem  in the eFuC case is given by:
\begin{equation}
\mathcal{P}_{\mathrm{eFuC},i}^{(\mathrm{A)}}\,:\underset{\left\Vert \underbar{\ensuremath{\bm{a}}}\right\Vert =1}{\mathrm{maximize}}\;\frac{\left(\underbar{\ensuremath{\bm{a}}}^{\dagger}\left[\underline{\bm{H}}^{e}(\bm{\Theta}_{\mathrm{fix}})\,\bm{D}_{\alpha}\,\ensuremath{\bar{\bm{\rho}}_{10}}\right]\right)^{2}}{\underbar{\ensuremath{\bm{a}}}^{\dagger}\overline{\mathrm{Cov}}_{\bm{\Theta}_{\mathrm{fix}}}(\text{\ensuremath{\underbar{\ensuremath{\bm{y}}}}}|\mathcal{H}_{i})\,\underbar{\ensuremath{\bm{a}}}}\,,
\end{equation}
where $\overline{\mathrm{Cov}}_{\bm{\Theta}_{\mathrm{fix}}}(\text{\ensuremath{\underbar{\ensuremath{\bm{y}}}}}|\mathcal{H}_{i})$
means that $\bm{\Theta}\rightarrow\bm{\Theta}_{\mathrm{fix}}$ in Eq.~\eqref{eq: augmented_covariance_mFuC}.
The optimal value of $\underbar{\ensuremath{\bm{a}}}$ in $\mathcal{P}_{\mathrm{eFuC},i}^{(\mathrm{A)}}$ is the vector attaining the equality in the \emph{Cauchy-Schwarz inequality}~\cite{Ciuonzo2025IoT}:
\begin{equation}
\text{\ensuremath{\underbar{\ensuremath{\bm{a}}}}}_{\mathrm{e,}i}^{\star}=\frac{\overline{\mathrm{Cov}}_{\bm{\Theta}_{\mathrm{fix}}}(\text{\ensuremath{\underbar{\ensuremath{\bm{y}}}}}|\mathcal{H}_{i})^{-1}\,\underline{\bm{H}}^{\mathrm{e}}(\bm{\Theta}_{\mathrm{fix}})\,\bm{D}_{\alpha}\,\bar{\bm{\rho}}_{10}}{\left\Vert \overline{\mathrm{Cov}}_{\bm{\Theta}_{\mathrm{fix}}}(\text{\ensuremath{\underbar{\ensuremath{\bm{y}}}}}|\mathcal{H}_{i})^{-1}\,\underline{\bm{H}}^{\mathrm{e}}(\bm{\Theta}_{\mathrm{fix}})\,\bm{D}_{\alpha}\,\bar{\bm{\rho}}_{10}\right\Vert }\,.\label{eq: mFuC fusion beamformer Step A - FC}
\end{equation}
Conversely, for bFuC case, the design problem for the set of WL vectors is:
\begin{align}
\mathcal{P}_{\mathrm{bFuC},i}^{(\mathrm{A)}}\, & :\underset{\left\Vert \underbar{\ensuremath{\bm{a}}}_{j}\right\Vert =1}{\mathrm{maximize}}\;\frac{\left(\underbar{\ensuremath{\bm{a}}}_{j}^{\dagger}\left[\underline{\bm{H}}^{e}(\bm{\Theta}_{\mathrm{fix}})\,\bm{D}_{\alpha}\,\ensuremath{\bm{\rho}_{10}}(\bm{p}^{\mathrm{t}}[j])\right]\right)^{2}}{\underbar{\ensuremath{\bm{a}}}_{j}^{\dagger}\,\mathrm{Cov}_{\bm{\Theta}_{\mathrm{fix}}}(\text{\ensuremath{\underbar{\ensuremath{\bm{y}}}}}|\mathcal{H}_{i};\bm{p}^{\mathrm{t}}[j])\,\underbar{\ensuremath{\bm{a}}}_{j}}\,\,.\nonumber \\
 & \quad j=1,\ldots N_{\mathrm{t}}
\end{align}
The optimal values of $\mathcal{P}_{\mathrm{bFuC},i}^{(\mathrm{A)}}$ are similarly obtained leveraging the Cauchy-Schwarz inequality~\cite{Ciuonzo2025IoT}:
\begin{align}
\text{\ensuremath{\underbar{\ensuremath{\bm{a}}}}}_{\mathrm{b},ij}^{\star} & =\frac{\mathrm{Cov}_{\bm{\Theta}_{\mathrm{fix}}}(\text{\ensuremath{\underbar{\ensuremath{\bm{y}}}}}|\mathcal{H}_{i};\bm{p}^{\mathrm{t}}[j])^{-1}\,\underline{\bm{H}}^{\mathrm{e}}(\bm{\Theta}_{\mathrm{fix}})\,\bm{D}_{\alpha}\,\ensuremath{\bm{\rho}_{10}}(\bm{p}^{\mathrm{t}}[j])}{\left\Vert \mathrm{Cov}_{\bm{\Theta}_{\mathrm{fix}}}(\text{\ensuremath{\underbar{\ensuremath{\bm{y}}}}}|\mathcal{H}_{i};\bm{p}^{\mathrm{t}}[j])^{-1}\,\underline{\bm{H}}^{\mathrm{e}}(\bm{\Theta}_{\mathrm{fix}})\,\bm{D}_{\alpha}\,\ensuremath{\bm{\rho}_{10}}(\bm{p}^{\mathrm{t}}[j])\right\Vert }\,.\label{eq: bFuC fusion beamformer Step A - FC}
\end{align}

\noindent
\textbf{Step (B) - RHS design:} We then focus on the optimization of the matrix $\bm{\Theta}$ for a \emph{fixed} WL vector  $\text{\ensuremath{\underbar{\ensuremath{\bm{a}}}}}_{\mathrm{fix}}$. 
The corresponding optimization problem is given by:
\begin{equation}
\mathcal{P}_{\mathrm{eFuC},i}^{(\mathrm{B)}}\,:\begin{array}{c}
\underset{\bm{\Theta}}{\mathrm{maximize}}\frac{\left(\underbar{\ensuremath{\bm{a}}}_{\mathrm{fix}}^{\dagger}\left[\underline{\bm{H}}^{e}(\bm{\Theta})\,\bm{D}_{\alpha}\,\ensuremath{\bar{\bm{\rho}}_{10}}\right]\right)^{2}}{\underbar{\ensuremath{\bm{a}}}_{\mathrm{fix}}^{\dagger}\overline{\mathrm{Cov}}_{\bm{\Theta}}(\text{\ensuremath{\underbar{\ensuremath{\bm{y}}}}}|\mathcal{H}_{i})\,\underbar{\ensuremath{\bm{a}}}_{\mathrm{fix}}}\\
\mathrm{subject\,to}\;\bm{\Theta}=\mathrm{diag}(e^{j\varphi_{1}},\ldots,e^{j\varphi_{M}})
\end{array}\,\,;
\end{equation}
\begin{equation}
\mathcal{P}_{\mathrm{bFuC},i}^{(\mathrm{B)}}\,:\begin{array}{c}
\underset{\bm{\Theta}}{\mathrm{maximize}}\sum_{j=1}^{N_{\mathrm{t}}}\frac{\left(\underbar{\ensuremath{\bm{a}}}_{j,\mathrm{fix}}^{\dagger}\left[\underline{\bm{H}}^{e}(\bm{\Theta})\,\bm{D}_{\alpha}\,\bm{\rho}_{10}(\bm{p}^{\mathrm{t}}[j])\right]\right)^{2}}{\underbar{\ensuremath{\bm{a}}}_{j,\mathrm{fix}}^{\dagger}\mathrm{Cov}_{\bm{\Theta}}(\text{\ensuremath{\underbar{\ensuremath{\bm{y}}}}}|\mathcal{H}_{i};\bm{p}^{\mathrm{t}}[j])\,\underbar{\ensuremath{\bm{a}}}_{j,\mathrm{fix}}}\\
\mathrm{subject\,to}\;\bm{\Theta}=\mathrm{diag}(e^{j\varphi_{1}},\ldots,e^{j\varphi_{M}})
\end{array}\,.
\end{equation}
After some manipulations, the deflection objectives can be recast as follows (isolating dependence on RHS phase-shifts):
\begin{equation}
g_{\mathrm{eFuC},i}(\bm{\theta})\triangleq\frac{\underbar{\ensuremath{\bm{\theta}}}^{\dagger}\bar{\bm{\Xi}}\left(\text{\ensuremath{\underbar{\ensuremath{\bm{a}}}}}_{\mathrm{fix}}\right)\underbar{\ensuremath{\bm{\theta}}}}{\underbar{\ensuremath{\bm{\theta}}}^{\dagger}\bar{\bm{\Psi}}_{i}\left(\text{\ensuremath{\underbar{\ensuremath{\bm{a}}}}}_{\mathrm{fix}}\right)\underbar{\ensuremath{\bm{\theta}}}}\,;\label{eq: D_mFuC_i}
\end{equation}
\begin{equation}
g_{\mathrm{bFuC},i}(\bm{\theta})\triangleq\sum_{j=1}^{N_{t}}\,\frac{\underbar{\ensuremath{\bm{\theta}}}^{\dagger}\bm{\Xi}_{j}\left(\text{\ensuremath{\underbar{\ensuremath{\bm{a}}}}}_{j,\mathrm{fix}}\right)\underbar{\ensuremath{\bm{\theta}}}}{\underbar{\ensuremath{\bm{\theta}}}^{\dagger}\bm{\Psi}_{ij}\left(\text{\ensuremath{\underbar{\ensuremath{\bm{a}}}}}_{j,\mathrm{fix}}\right)\underbar{\ensuremath{\bm{\theta}}}}\,.\label{eq: D_bFuC_i}
\end{equation}
In what follows, for notational compactness, we drop the dependence of 
$\bar{\bm{\Xi}}$ and $\bar{\bm{\Psi}}_{i}$ on $\text{\ensuremath{\underbar{\ensuremath{\bm{a}}}}}_{\mathrm{fix}}$
(resp. $\bm{\Xi}_{j}$ and $\bm{\Psi}_{ij}$ on $\text{\ensuremath{\underbar{\ensuremath{\bm{a}}}}}_{j,\mathrm{fix}}$).

In the above expression, the matrix $\bar{\bm{\Xi}}$ is defined as $\bar{\bm{\Xi}}\triangleq\left(\bar{\bm{N}}^{\dagger}\text{\ensuremath{\underbar{\ensuremath{\bm{a}}}}}_{\mathrm{fix}}\right)\left(\bar{\bm{N}}^{\dagger}\text{\ensuremath{\underbar{\ensuremath{\bm{a}}}}}_{\mathrm{fix}}\right)^{\dagger}$
where the auxiliary matrix term $\bar{\bm{N}}$ has the following explicit expression:
\begin{equation}
\bar{\bm{N}}\triangleq\begin{bmatrix}\bm{G}\,\mathrm{diag}(\bm{H}\,\bm{D}_{\alpha}\,\bar{\bm{\rho}}_{10}) & \bm{O}_{N\times M}\\
\bm{O}_{N\times M} & \bm{G}^{\ast}\,\mathrm{diag}^{\ast}(\bm{H}\,\bm{D}_{\alpha}\,\bar{\bm{\rho}}_{10})
\end{bmatrix}\,.\label{eq: ENNE}
\end{equation}
Similarly, the matrices $\bm{\Xi}_{j}$ are defined as $\bm{\Xi}_{j}\triangleq\left(\bm{N}_{j}^{\dagger}\text{\ensuremath{\underbar{\ensuremath{\bm{a}}}}}_{j,\mathrm{fix}}\right)\left(\bm{N}_{j}^{\dagger}\text{\ensuremath{\underbar{\ensuremath{\bm{a}}}}}_{j,\mathrm{fix}}\right)^{\dagger}$, where $\bm{N}_{j}$ has a similar expression as Eq. (34) when replacing $\bar{\bm{\rho}}_{10}$ with $\bm{\rho}_{10}(\bm{p}^{\mathrm{t}}[j])$.

Differently, the matrix $\bar{\bm{\Psi}}_{i}$ at the denominator of Eq.~\eqref{eq: D_mFuC_i} is defined as follows:
\begin{gather}
\bar{\bm{\Psi}}_{i}\triangleq\bm{\Delta}_{0}^{\dagger}\,\bm{D}_{\alpha}\,\overline{\mathrm{Cov}}(\bm{x}|\mathcal{H}_{i})\,\bm{D}_{\alpha}\,\bm{\Delta}_{0}+\frac{\sigma_{w}^{2}}{2M}\,\left\Vert \text{\ensuremath{\underbar{\ensuremath{\bm{a}}}}}_{\mathrm{fix}}\right\Vert ^{2}\bm{I}_{2M}\,,\label{eq: avg_Psi}
\end{gather}
where
\begin{equation}
\bm{\Delta}_{0}\triangleq\begin{bmatrix}\bm{H}^{T}\,\mathrm{diag}^{\ast}(\bm{G}^{\dagger}\,\bm{a}_{\mathrm{fix}}) & \bm{H}^{\dagger}\,\mathrm{diag}(\bm{G}^{\dagger}\,\bm{a}_{\mathrm{fix}})\end{bmatrix}\,.
\end{equation}
Similarly, the matrices $\bm{\Psi}_{ij}$ in Eq.~\eqref{eq: D_bFuC_i} are obtained by replacing $\overline{\mathrm{Cov}}(\bm{x}|\mathcal{H}_{i})$ with $\mathrm{Cov}(\bm{x}|\mathcal{H}_{i};\bm{p}^{\mathrm{t}}[j])$ in Eq.~\eqref{eq: avg_Psi}, as well as by replacing $\bm{a}_{\mathrm{fix}}$ with $\bm{a}_{j,\mathrm{fix}}$ in $\bm{\Delta}_{0}$.
Despite having rewritten the objectives in Eqs. \eqref{eq: D_mFuC_i} and \eqref{eq: D_bFuC_i} in a more explicit form, due to modulus constraint of RHS elements and because the objective is a ratio (resp. sum of ratios) of quadratic forms with positive semidefinite matrices, the problem is \emph{non-convex} in both cases.
Hence, to avoid cumbersome optimizations at the FC, we propose to solve $\mathcal{P}_{\mathrm{eFuC},i}^{(\mathrm{B)}}$ and $\mathcal{P}_{\mathrm{bFuC},i}^{(\mathrm{B)}}$ 
via the MM technique~\cite{sun2016majorization}. 
This is accomplished as follows.

Let $\bm{\theta}_{(\ell)}^{\star}$ denote the value of the phase-shift vector at the $\ell$th AO iteration.
At each step, we construct a surrogate function $f(\bm{\theta} | \bm{\theta}_{(\ell)}^{\star})$ that lower-bounds the original objective $g(\bm{\theta})$ and is tight at $\bm{\theta} = \bm{\theta}_{(\ell)}^{\star}$.
This surrogate is then maximized (with the same constraints as the original objective) to obtain the next iterate, $\bm{\theta}_{(\ell+1)}^{\star}$, thereby ensuring that the objective does not decrease across iterations, i.e., $g(\bm{\theta}_{(\ell+1)}^{\star}) \geq g(\bm{\theta}_{(\ell)}^{\star})$. This guarantees first-order optimality of the resulting solution.
The success of this MM strategy hinges on designing a surrogate $f(\bm{\theta} | \bm{\theta}_{(\ell)}^{\star})$ that is both tractable and leads to a closed-form (or efficiently computable) update for the solution of the surrogate optimization problem $\ensuremath{\breve{\mathcal{P}}^{(\mathrm{B)}}}$, denoted with $\bm{\theta}_{(\ell+1)}^{\star}$. 
The following lemma provides closed-form updates associated with optimization of surrogates for the RHS optimization sub-problems $\bar{\mathcal{P}}_{\mathrm{eFuC},i}^{(\mathrm{B)}}$ and $\bar{\mathcal{P}}_{\mathrm{bFuC},i}^{(\mathrm{B)}}$.

\begin{lemma}
The optimization problems $\breve{\mathcal{P}}{}_{\mathrm{eFuC,}i}^{(\mathrm{B)}}$ and $\breve{\mathcal{P}}{}_{\mathrm{bFuC,}i}^{(\mathrm{B)}}$ admit closed-form solutions, which can be compactly expressed as follows.
For the bFuC-based design, the optimal phase vector at iteration $(\ell+1)$ is given by:
\begin{gather}
\angle\text{\ensuremath{\underbar{\ensuremath{\bm{\theta}}}}}_{(\ell+1)}^{\star}=\angle\left(\sum_{j=1}^{N_{\mathrm{t}}}\frac{\bm{\Xi}_{j}\,\underbar{\ensuremath{\bm{\theta}}}_{(\ell)}^{\star}}{\left(\underbar{\ensuremath{\bm{\theta}}}_{(\ell)}^{\star}\right)^{\dagger}\bm{\Psi}_{ij}\,\underbar{\ensuremath{\bm{\theta}}}_{(\ell)}^{\star}}-\right.\label{eq: RHS optimization (Step 2) fc}\\
\left.\sum_{j=1}^{N_{\mathrm{t}}}\frac{\left(\underbar{\ensuremath{\bm{\theta}}}_{(\ell)}^{\star}\right)^{\dagger}\,\bm{\Xi}_{j}\,\underbar{\ensuremath{\bm{\theta}}}_{(\ell)}^{\star}}{\left(\left(\underbar{\ensuremath{\bm{\theta}}}_{(\ell)}^{\star}\right)^{\dagger}\bm{\Psi}_{ij}\,\underbar{\ensuremath{\bm{\theta}}}_{(\ell)}^{\star}\right)^{2}}\,\left(\bm{\Psi}_{ij}-\lambda_{max}(\bm{\Psi}_{ij})\,\bm{I}_{2M}\right)\,\underbar{\ensuremath{\bm{\theta}}}_{(\ell)}^{\star}\right)\,.\nonumber 
\end{gather}
In the case of eFuC-based design, the same expression applies with the substitutions $N_{\mathrm{t}}=1$, $\bm{\Xi}_{j}\rightarrow\bar{\bm{\Xi}}$, and $\bm{\Psi}_{ij}\rightarrow\bar{\bm{\Psi}}_{i}$.
\end{lemma}

\begin{proof}
The proof is given in the Appendix.
\end{proof}

\subsection{Summary and Complexity Overview} \label{subsec:complexity}

The designed eFuC and bFuC procedures are then summarized in \textbf{Algorithms~\ref{alg:AO-MM-mFuC}} and \textbf{\ref{alg:AO-MM-bFuC}}.
In both cases the resulting design alternates between the closed-form updates related to fusion rule [\textbf{Step (A)}] and RHS design [\textbf{Step (B)}].
Owing to the monotonic ascent of the deflection objective across iterations, the proposed AO procedure ensures convergence in the objective value. As a result, the selected deflection metric is guaranteed to monotonically increase and converge to a \emph{local optimum}.
For simplicity, the initial point $\bm{\theta}_{(0)}^{\star}$ is obtained using uniformly generated random phase shifts. Nonetheless, more structured initialization strategies can be readily incorporated if desired.
\begin{algorithm}[htbp]\caption{eFuC Design via Alternating Optimization.}\label{alg:AO-MM-mFuC}
\begin{mdframed}[backgroundcolor=lightbluebox!45, linecolor=lightbluebox!65, linewidth=1pt, shadow=true,
shadowsize=2pt,
shadowcolor=lightbluebox!95]
\begin{algorithmic}[1]
\STATE  Initialize the vector $\bm{\theta}_{(0)}^{\star}$ and set $\ell=0$;
\REPEAT
\STATE For fixed $\bm{\theta}_{(\ell)}^{\star}$, update $\underbar{\ensuremath{\bm{a}}}$ via Eq.~\eqref{eq: mFuC fusion beamformer Step A - FC};
\STATE For fixed $\underbar{\ensuremath{\bm{a}}}$, update $\bm{\theta}_{(\ell+1)}^{\star}$ via Eq.~\eqref{eq: RHS optimization (Step 2) fc};
\STATE Increment index: $\ell \gets \ell + 1$;
\UNTIL convergence criterion is met 
\end{algorithmic}
\end{mdframed}
\end{algorithm}

\begin{algorithm}[htbp]\caption{bFuC Design via Alternating Optimization.}\label{alg:AO-MM-bFuC}
\begin{mdframed}[backgroundcolor=lightpurplebox!45, linecolor=lightpurplebox!65, linewidth=1pt, shadow=true,
shadowsize=2pt,
shadowcolor=lightpurplebox!95]
\begin{algorithmic}[1]
\STATE  Initialize the vector $\bm{\theta}_{(0)}^{\star}$ and set $\ell=0$;
\REPEAT
\STATE For fixed $\bm{\theta}_{(\ell)}^{\star}$, update $\left\{ \underbar{\ensuremath{\bm{a}}}_{j}\right\} _{j=1}^{N_{\mathrm{t}}}$ via Eq.~\eqref{eq: bFuC fusion beamformer Step A - FC};
\STATE For fixed $\left\{ \underbar{\ensuremath{\bm{a}}}_{j}\right\} _{j=1}^{N_{\mathrm{t}}}$, update $\bm{\theta}_{(\ell+1)}^{\star}$ via Eq.~\eqref{eq: RHS optimization (Step 2) fc};
\STATE Increment index: $\ell \gets \ell + 1$;
\UNTIL convergence criterion is met 
\end{algorithmic}
\end{mdframed}
\end{algorithm}

The \emph{overall complexity} of the two proposed (AO-based) joint design approaches (eFuC and bFuC) is $\mathcal{O}(N_{\mathrm{iter}}\,(C_{\mathrm{fus}}+C_{\mathrm{rhs}} ) + C_{\mathrm{init}})$, where  $N_{\mathrm{iter}}$ is the number of outer iterations, and $C_{\mathrm{fus}}$, $C_{\mathrm{rhs}}$ denote the costs of the fusion vector update and RHS phase optimization, respectively. The initialization cost $C_{\mathrm{init}}$ refers to operations performed once, outside the AO loop.
All costs are summarized in Tab.~\ref{tab: Complexity comparison}.
For completeness, the complexity of the IS-based design from~\cite{Ciuonzo2025IoT} is also included in the same table.
Clearly, the IS-based design incurs \emph{lower complexity} because it ignores the statistical characteristics of the sensing process (i.e. naively assuming sensors with $(P_{\mathrm{d},k},P_{\mathrm{f},k})=(1,0)$ $\forall k\in\mathcal{K}$).
As a result, its detection performance is expected to degrade when sensors deviate from this idealized behavior.

Regarding \textbf{Step (A)}, the complexity of eFuC-based design is mainly given by the matrix products needed to obtain $\bm{H}^{e}(\bm{\Theta})$ and the augmented covariance $\overline{\mathrm{Cov}}_{\bm{\Theta}_{\mathrm{fix}}}(\text{\ensuremath{\underbar{\ensuremath{\bm{y}}}}}|\mathcal{H}_{i})$ (cf. Eq.~\eqref{eq: augmented_covariance_mFuC}), as well as to invert the latter matrix.
Conversely, for bFuC-based design, most of the above computations scale with the number of filters $N_{\mathrm{t}}$, save from the computation of $\bm{H}^{e}(\bm{\Theta})$.
Still, for bFuC-0 there is a lower complexity w.r.t. bFuC-1, since the covariance $\mathrm{Cov}_{\bm{\Theta}_{\mathrm{fix}}}(\text{\ensuremath{\underbar{\ensuremath{\bm{y}}}}}|\mathcal{H}_{0})$ does not depend on the target position. Hence, this need to be computed and inverted once.
Regarding \textbf{Step (B)}, the complexity of eFuC-based design is dominated by the computation of $\bm{\Delta}_{0}$, evaluating 
$\bar{\bm{\Psi}}_{i}$ and determining  $\lambda_{max}(\bar{\bm{\Psi}}_{i})$.
On the contrary for bFuC-based design, \emph{all these computations} need to be executed separately for the $N_{\mathrm{t}}$ filters.

Once the design procedure has been carried out, it is worth noticing that given the WL fusion assumption made in this work (cf. Eq.~\eqref{eq: WL fusion statistic}), the complexity of the DD task (see Tab.~\ref{tab: complexitypery}) at the FC is $\mathcal{O}(N)$ for \emph{both} IS and eFUC cases, since the RHS carries out analog processing.
The advantage of eFuC over the IS counterpart stems from its use of a WL fusion vector that is \emph{expressly optimized to account for the sensing process and its associated uncertainties} (cf. Eqs.~\eqref{eq: deflection_mFuC1} and \eqref{eq: deflection_mFuC0}), rather than ignoring them as in IS-based design.
On the contrary, for bFuC-based design, the digital part of the fusion rule implementation is based on a filter-bank of WL rules (cf. Eq.~\eqref{eq: FB-WL fusion statistic}). Hence the complexity in the latter case is $\mathcal{O}(NN_{\mathrm{t}})$.

The complexity of the latter rules stands in sharp contrast to that of a GLR-based implementation, which requires $\mathcal{O}\left(2^{K}(N+N_{t})\right)$ operations (cf. Eq.~\eqref{eq:GLR_RIS}).
It is important to note that, although the GLR also scales linearly with the grid size $N_{t}$--similar to bFuC--the dominant term in its complexity arises from the exponential dependence on the number of sensors $K$.
In contrast, the bFuC design is able to effectively exploit implicit localization through grid discretization while ($a$) maintaining linear complexity in $N_{t}$ and ($b$) avoiding the exponential growth with $K$ that makes GLR infeasible in practical WSN deployments.

\renewcommand{\arraystretch}{1.7}
\begin{table*}
\centering
\caption{
Computational complexity summary for the joint eFuC and bFuC design strategies. The complexity of the joint IS design is included for comparison. For each approach, the fusion step ($C_{\mathrm{fus}}$), RHS design step ($C_{\mathrm{rhs}}$), and initialization outside the AO loop ($C_{\mathrm{init}}$) are highlighted. Here, $K$ is the number of sensors, $M$ the number of RHS elements, $N$ the number of receive feeds, and $N_{\mathrm{t}}$ the number of target position grid points.\label{tab: Complexity comparison}}
\begin{tabular}{c c c c}
\hline 
\multirow{2}{*}{\shortstack{\textbf{Design} \\ \textbf{Principle}}} & \multicolumn{3}{c}{\textbf{Complexity}} \\
\cline{2-4}
& \textbf{Fusion Rule Update ($C_{\mathrm{fus}}$)} & \textbf{RHS Update ($C_{\mathrm{rhs}}$)} & \textbf{Init ($C_{\mathrm{init}}$)} \\
\hline 
\rowcolor[HTML]{D6D6FF} eFuC & $\mathcal{O}(M^2K + NMK + K^2N + KN^2 + N^3)$ & $\mathcal{O}(M^{3} + MN + K^{2}M + M^{2}K + KN)$ & -- \\
\rowcolor[HTML]{F0C8F0} bFuC-0 & $\mathcal{O}(M^{2}K+NMK+K^{2}N+(N+N_{\mathrm{t}})(KN+N^{2}))$ & $\mathcal{O}(N_{\mathrm{t}}(M^{3}+MN+K^{2}M+M^{2}K+KN))$ & -- \\
\rowcolor[HTML]{F0C8F0} bFuC-1 & $\mathcal{O}(M^{2}K+NMK+N_{\mathrm{t}}(K^{2}N+KN^{2}+N^{3}))$ & $\mathcal{O}(N_{\mathrm{t}}(M^{3}+MN+K^{2}M+M^{2}K+KN))$ & -- \\
\rowcolor[HTML]{f8e8e8} IS & $\mathcal{O}(MN)$ & $\mathcal{O}(M^{2} + MN)$ & $\mathcal{O}(NK + MK + NM)$ \\
\hline 
\end{tabular}
\end{table*}
\renewcommand{\arraystretch}{1}

\renewcommand{\arraystretch}{1.7}
\begin{table}
\begin{centering}
\medskip{}
\par\end{centering}
\centering{}\caption{ Operational complexity of the joint eFuC and bFuC design strategies. For comparison, the complexity of the IS-based and GLR-based detection strategies is also reported. \label{tab: complexitypery}}
\begin{tabular}{cccc}
\hline 
\textbf{Fusion Rule} & \textbf{Complexity (for each $\bm{y})$}\tabularnewline
\hline 
\rowcolor[HTML]{DCDCDC}GLR & $\mathcal{O}(2^K(N+N_{\mathrm{t}}))$ \tabularnewline
\rowcolor[HTML]{D6D6FF}eFuC & $\mathcal{O}(N)$  \tabularnewline
\rowcolor[HTML]{F0C8F0}bFuC & $\mathcal{O}(NN_{\mathrm{t}})$  \tabularnewline
\rowcolor[HTML]{f8e8e8}IS & $\mathcal{O}(N)$  \tabularnewline
\hline 
\end{tabular}
\end{table}
\renewcommand{\arraystretch}{1}

\section{Simulation Results} \label{Sim_res}
\subsection{Simulated RHS-WSN-PoI setup}

\noindent 
\textbf{WSN-RHS-FC displacement:} All spatial quantities are expressed in units of the wavelength $\lambda$. Sensor nodes are randomly distributed in $[0,40]\times[0,40]\times[0,3]\,\lambda$. The RHS is a square planar array centered at $[70,20,10]\,\lambda$, aligned along the y-axis, with element spacing $\lambda/3$. 
The $N$ (external) receive feeds are arranged in a uniformly spaced linear (or planar) array, oriented parallel to the x-axis, with inter-element distance equal to $\lambda/2$, and geometrically centered at the spatial coordinate $[68, 18, 10]\,\lambda$.
Without loss of generality, we assume that the element side lengths coincide with the corresponding element spacings. 
As a result, it holds $\Delta_{\mathrm{h}}^{\mathrm{rhs}}=\Delta_{\mathrm{v}}^{\mathrm{rhs}}=\lambda/3$ and $\Delta_{\mathrm{h}}^{\mathrm{fc}}=\Delta_{\mathrm{v}}^{\mathrm{fc}}=\lambda/2$, respectively.

\noindent 
\textbf{Sensing parameters:} We model the sensing noise as $n_k \sim \mathcal{N}(0,1)$, $\forall k \in \mathcal{K}$, and adopt a power-law AAF $g(\bm{p}^{\mathrm{t}},\bm{p}_{k}^{\mathrm{sen}})\triangleq1\,/\,\sqrt{1+\left(\left\Vert \bm{p}^{\mathrm{t}}-\bm{p}_{k}^{\mathrm{sen}}\right\Vert /\,\eta\right)^{\alpha}}$
with $\eta = 12 \lambda$, $\alpha = 4$. The target SNR is defined as $\mathrm{SNR}_{\mathrm{t}} = 10\log_{10}(\theta/\sigma_w^2)$, set to $15\,\mathrm{dB}$.

\noindent 
\textbf{Channel parameters:} Sensors transmit with equal power $\alpha_k = 1$, and experience a path loss with reference gain $\mu = -30\,\mathrm{dB}$ at $d_0 = 1\,\lambda$, and exponent $\nu = 2$. The (randomly-drawn) Rician factor $\kappa_k \in (3,5)\,\mathrm{dB}$ yields $b_k = \sqrt{ \kappa_k / (1 + \kappa_k) }$.
The directivity gain in Eq.~\eqref{eq: general directivity pattern} is shaped via $q$ to emulate a $\cos^3(\cdot)$ pattern, achieving a peak gain of $8$ ($\approx 9.03\,\mathrm{dBi}$). RHS reflection efficiency is set to $\eta = 1$, and unless stated otherwise, (channel) noise variance is $\sigma_w^2 = -50\,\mathrm{dBm}$.
  
\noindent 
\textbf{Simulation setup:} Each DD result reported in what follows is averaged over: ($i$) $10^2$ independent channel realizations of $\bm{H}$ (since $\bm{G}$ is deterministic given the chosen RHS-feed configuration); ($ii$) $10^3$ Monte Carlo trials per realization, with target randomly located within the square area on the ground $\mathcal{A}\triangleq[0,40]\times[0,40]\times[0]\lambda$.
Following Sec.~\ref{subsec: WLlinear_design}, both $\Lambda_{\mathrm{GLR}}$ and $\Lambda_{\mathrm{bWL}}$ are evaluated via grid discretization of the surveillance area $\mathcal{A}$.
When not otherwise specified, the surveillance area is uniformly sampled with $N_t = N_c^2$, where $N_c = 5$.

\subsection{Upper bound and baselines considered}

In the following analysis, the \emph{observation bound} is also reported for completeness (curve \say{GLR Obs. bound} in the following figures). 
The latter represents the performance obtained by a GLR which operates over an \emph{ideal channel condition} (i.e. all the $K$ sensors deliver their decisions noise-free to the FC):
\begin{gather}
\Lambda_{\mathrm{GLR}}^{\mathrm{obs}}=\max_{\bm{p}^{\mathrm{t}}}\ln\left[\frac{f_{1}(\bm{x};\bm{p}^{\mathrm{t}})}{f_{0}(\bm{x})}\right]=\max_{\bm{p}^{\mathrm{t}}}\left\{ \sum_{k=1}^{K}\frac{(1+x_{k})}{2}\times\right.\label{eq:GLRobs_RHS}\\
\left.\ln\left[\frac{P_{\mathrm{d},k}(\bm{p}^{\mathrm{t}})}{P_{\mathrm{f},k}}\right]+\frac{(1-x_{k})}{2}\ln\left[\frac{1-P_{\mathrm{d},k}(\bm{p}^{\mathrm{t}})}{1-P_{\mathrm{f},k}}\right]\right\} \,.\nonumber 
\end{gather}
The above bound represents a relevant benchmark to assess both ($i$) the detection degradation due to the interfering distributed MIMO channel and ($ii$) the corresponding benefit arising from the RHS adoption.
$\Lambda_{\mathrm{GLR}}^{\mathrm{obs}}$ is implemented similarly as $\Lambda_{\mathrm{GLR}}$ and $\Lambda_{\mathrm{bWL}}$ via grid discretization in what follows.

Secondly, as a \emph{baseline}, we consider the joint IS design considered in \cite{mudkey2022wireless,Ciuonzo2025IoT}.
Specifically, the aforementioned design is based on the ideal-sensors assumption, namely $(P_{\mathrm{d},k},P_{\mathrm{f},k}) = (1,0), \forall k$. 
Hence, this design is agnostic of sensor performance and can be implemented without knowledge of target position 
$\bm{p}^{\mathrm{t}}$.
The design operates in an iterative fashion between the two steps. Namely, the fusion rule step equals:
\begin{equation}
\text{\ensuremath{\underbar{\ensuremath{\bm{a}}}}}_{\mathrm{IS}}^{\star}(\bm{\Theta}_{\mathrm{fix}})=\frac{\underline{\bm{H}}^{\mathrm{e}}(\bm{\Theta}_{\mathrm{fix}})\,\bm{D}_{\alpha}\,\bm{1}_{k}}{\left\Vert \underline{\bm{H}}^{\mathrm{e}}(\bm{\Theta}_{\mathrm{fix}})\,\bm{D}_{\alpha}\,\bm{1}_{k}\right\Vert }\,.\label{eq: WL fusion beamformer Step A - IS}
\end{equation}
Conversely, the RHS update step equals:
\begin{equation}
\angle\text{\ensuremath{\underbar{\ensuremath{\bm{\theta}}}}}_{(\ell+1)}^{\star}=\angle\left(\bm{\Xi}_{\mathrm{IS}}\,\text{\ensuremath{\underbar{\ensuremath{\bm{\theta}}}}}_{(\ell)}^{\star}\right)\,,\label{eq: RIS optimization (Step 2) is}
\end{equation}
where $\bm{\Xi}_{\mathrm{IS}}$
is given by $\bm{\Xi}_{\mathrm{IS}}\triangleq\left(\bm{N}_{\mathrm{IS}}^{\dagger}\,\text{\ensuremath{\underbar{\ensuremath{\bm{a}}}}}_{\mathrm{fix}}\right)\left(\bm{N}_{\mathrm{IS}}^{\dagger}\,\text{\ensuremath{\underbar{\ensuremath{\bm{a}}}}}_{\mathrm{fix}}\right)^{\dagger}$, and $\bm{N}_{\mathrm{IS}}$ is obtained by replacing $\bar{\bm{\rho}}_{10}\rightarrow\bm{1}_{K}$ into $\bar{\bm{N}}$ (cf. Eq.~\eqref{eq: ENNE}).

\subsection{Performance Gaps in Target Detection with Existing RHS-aided Fusion Rules}

\begin{figure}
\centering{}\includegraphics[width=1\columnwidth]{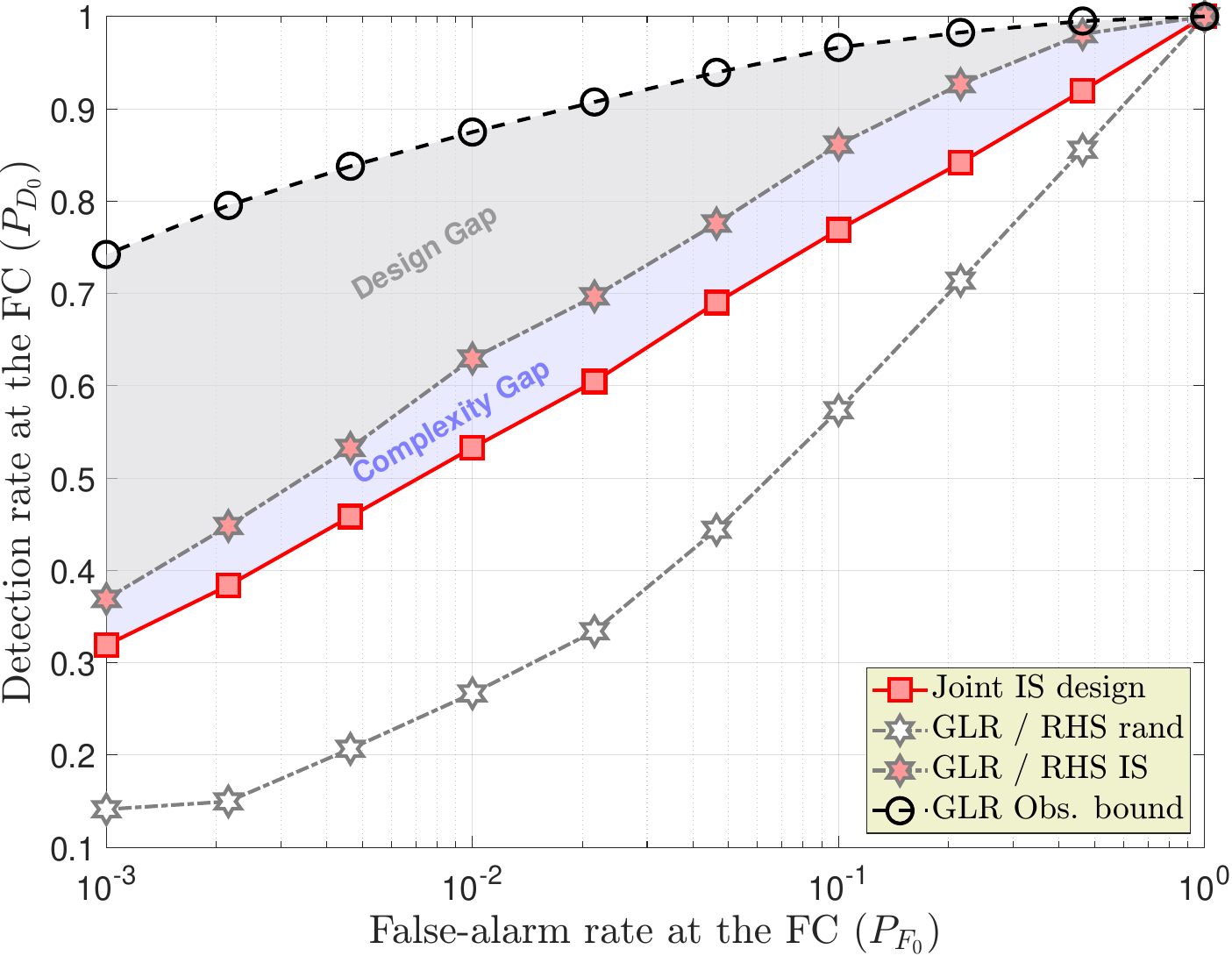}\caption{Assessing the gaps of existing joint design in holographic DF via ROCs ($P_{D_{0}}$ vs $P_{F_{0}}$). WSN with $K=15$ sensors and target emission strength set to $\mathrm{SNR_{sen}}=15\,\mathrm{dB}$.
Holographic DF is implemented with an RHS made of $M=64$ elements and $N=1$ receive feed; channel noise variance is set to $\sigma_{w}^{2}=-50$ dBm. \label{fig:ROC_gap}
}
\end{figure}

Fig.~\ref{fig:ROC_gap} illustrates the detection performance at the FC using Receiver Operating Characteristic (ROC) curves, 
where the detection probability $P_{D_{0}}$ is plotted against the false-alarm probability $P_{F_{0}}$, for
existing system design strategies.
A WSN with $K = 15$ sensors is considered, and the holographic receiver comprises $M = 64$ RHS elements and a single feed ($N = 1$).
The analysis highlights key limitations of current RHS architectures and fusion rules in the absence of RHS-specific optimization.
The \say{GLR Obs. bound} curve represents the ideal performance limit of GLR-based detection under channel error-free observationsm (cf. Eq.~\eqref{eq:GLRobs_RHS}).

Despite being both optimized over the RHS configuration and inference strategy, the joint IS design \emph{exhibits a substantial performance gap} relative to the theoretical limit. For instance, at $P_{F0} = 10^{-2}$, the gap in detection probability is approximately $30\%$. 
Nevertheless, the joint IS strategy outperforms the (conventional) GLR fusion rule (cf. Eq.~\eqref{eq:GLR_RIS}) applied to a randomly-configured RHS, thereby demonstrating that even partial exploitation of the RHS’s latent degrees of freedom can yield meaningful performance gains.

To further dissect complexity-performance tradeoffs, we also consider the  \say{GLR / RHS IS} approach, where analog RHS processing originates from the joint IS design, while the digital processing at the FC employs the GLR rule. This hybrid scheme improves performance compared to a fully-random RHS configuration, underscoring the value of analog-side optimization even when the target location is unknown (but ignored). However, this comes at a steep computational cost (\emph{complexity gap}): the GLR rule requires evaluating all $2^{15} \approx 32\text{k}$ decision vectors, rendering the approach practically infeasible.

Crucially, even this hybrid design fails to close the gap to the fundamental bound, exposing a large \emph{design gap} in existing approaches--the shortfall induced by suboptimal RHS configuration.
In conclusion, the results point to a critical need for novel joint RHS-fusion rule designs under partial knowledge of the sensing environment, particularly in scenarios involving unknown target locations.

\subsection{Assessing the Gain of the Proposed Solutions}

\begin{figure}
\centering{}\includegraphics[width=1\columnwidth]{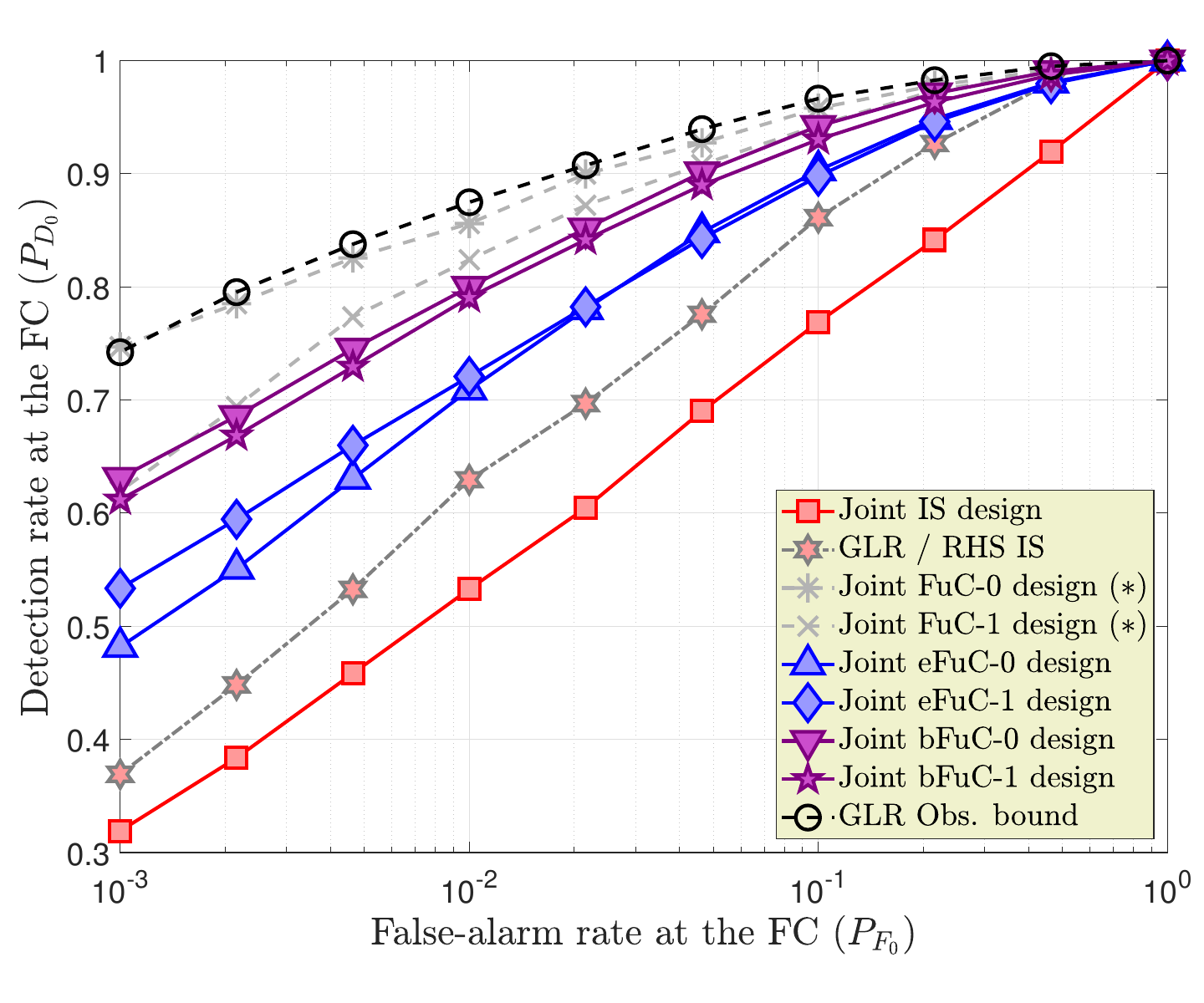}\caption{Assessing the gains of eFuC/bFuC joint design in holographic DF via ROCs ($P_{D_{0}}$ vs $P_{F_{0}}$). WSN with $K=15$ sensors and target emission strength set to $\mathrm{SNR_{sen}}=15\,\mathrm{dB}$. Holographic DF is implemented with an RHS made of $M=64$ elements and $N=1$ receive feed; channel noise variance is set to $\sigma_{w}^{2}=-50$ dBm. \label{fig:ROC_proposed}
}
\end{figure}

In Fig.~\ref{fig:ROC_proposed}, we provide a comparative performance assessment between the proposed eFuC- and bFuC-based approaches and the baseline joint IS design. A clear performance hierarchy emerges. Both joint design strategies \emph{substantially enhance detection capabilities}, underscoring the merits of integrated decision-making in the presence of uncertainty. Notably, the bFuC-based solution consistently outperforms its eFuC counterpart.
This is a direct consequence of the bFuC formulation, \emph{which incorporates the estimation of the unknown target position $\bm{p}^{\mathrm{t}}$ via the filter-bank structure in Eq.~\eqref{eq: FB-WL fusion statistic}} (as opposed to the simpler WL processing of eFuC design, see Eq.~\eqref{eq: WL fusion statistic}), thereby enabling a more informed fusion of observations.

To illustrate, for a nominal false alarm probability of $P_{F_0}=10^{-2}$, the joint IS design yields a detection probability marginally $\geq0.5$.
In contrast, the eFuC and bFuC schemes attain detection probabilities of approximately $0.7$ and $0.8$, respectively--\emph{substantially narrowing the performance gap to the GLR observation bound}.
The proposed solutions further surpass a computationally-intensive benchmark alternative based on GLR fusion with an IS-designed RHS matrix, which attains a detection probability of $\approx0.63$.
For completeness, and to provide an additional benchmark, we also report the performance of \emph{clairvoyant} FuC-based solutions~\cite{Ciuonzo2025IoT} under the idealized assumption of \emph{known target position} $\bm{p}^{\mathrm{t}}$ at each trial. These results serve as upper bounds, delineating the inherent challenge of handling positional uncertainty within WL-based fusion rule paradigm. Remarkably, the performance of the bFuC scheme remains relatively close to this ideal limit, highlighting its robustness and efficacy even under imperfect knowledge conditions.

When comparing designs optimized under different deflection measures (e.g., eFuC-0 vs. eFuC-1 and bFuC-0 vs. bFuC-1), a difference in ranking emerges. 
In the eFuC case, eFuC-1 consistently outperforms eFuC-0, likely because it exploits the covariance structure under $\mathcal{H}_1$, thus capturing additional information about the received signal $\bm{y}$.
Interestingly, this advantage does not carry over to the bFuC family. For bFuC, the filter-bank structure already provides strong implicit localization capability, and incorporating the $\mathcal{H}_1$ covariance within each filter does not yield further gains; in fact, it may even be slightly detrimental. 
This suggests that, for bFuC, the spatial filtering stage largely resolves the underlying uncertainty, reducing the marginal benefit of a target-covariance-aware metric.

\subsection{Detection Performance vs. Grid Size}
\begin{figure}
\centering{}\includegraphics[width=0.48\textwidth]{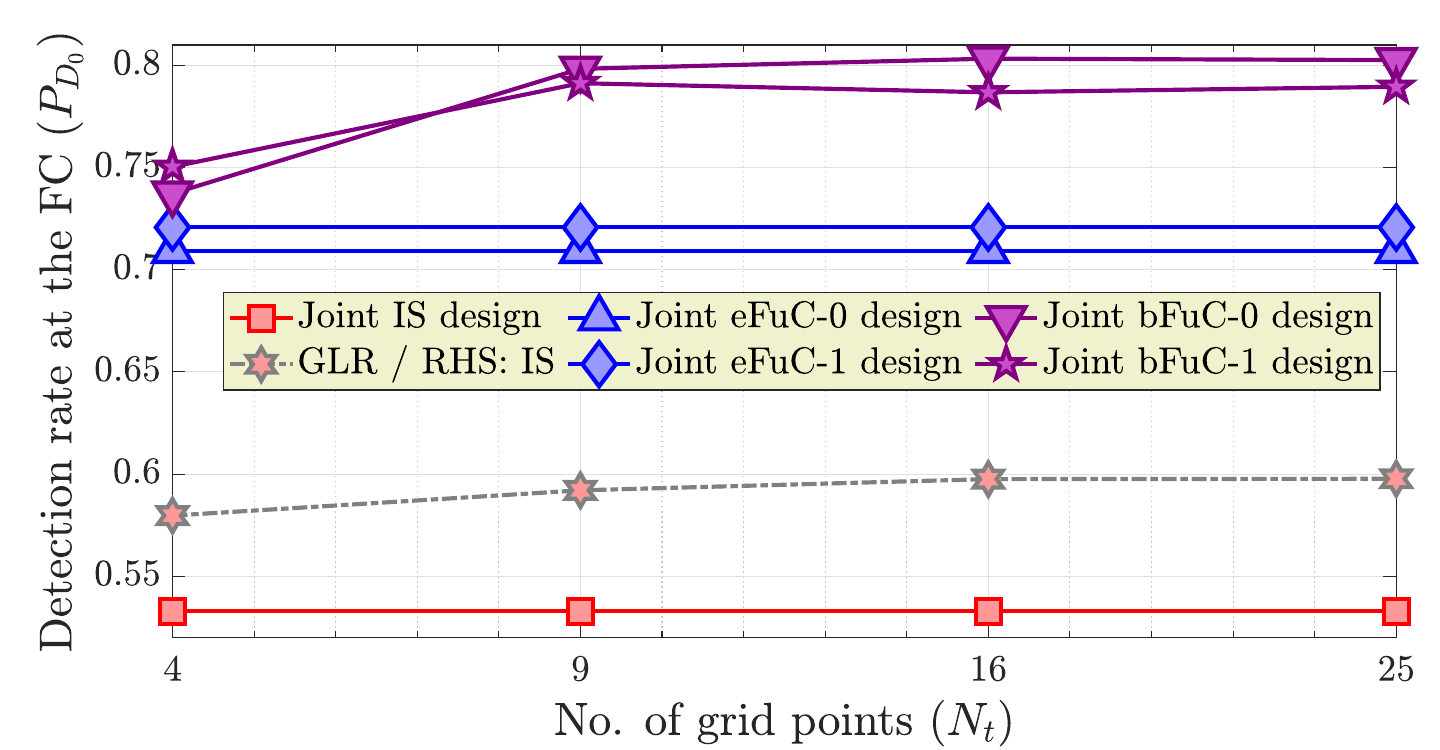}\caption{Assessing the sensitivity of bFuC joint design w.r.t. the grid size $N_t$ via $P_{D_{0}}$, where the false-alarm rate is set to $P_{F_{0}}=10^{-2}$. WSN with $K=15$ sensors and target emission strength set to $\mathrm{SNR_{sen}}=15\,\mathrm{dB}$. Holographic DF is implemented with an RHS made of $M=64$ elements and $N=1$ receive feed; channel noise variance is set to $\sigma_{w}^{2}=-50$ dBm.
\label{fig:Pd_vs_gridsize}}
\end{figure}

Given the dependence of bFuC approaches on the grid size $N_t$, Fig.~\ref{fig:Pd_vs_gridsize} illustrates their detection probability $P_{D_0}$ at a fixed false-alarm rate $P_{F_0} = 0.01$, as the grid size varies according to $N_{t}=N_{c}^{2}$ with $N_{c}\in \{2,3,4,5\}$.
As in the previous examples, a WSN with $K=15$ sensors is considered.
For completeness, we also report the performance of eFuC design solutions together with that of IS-based design (which are \emph{all} invariant to $N_t$).
In addition, the detection performance of the GLR when the RHS is obtained via the IS principle is shown as a function of $N_t$.

The results indicate a noticeable performance gap for grid sizes $N_t\in\{4,9\}$, whereas for $N_t=16$ and $N_t=25$ the performance of both bFuC variants, as well as that of the GLR with IS-based RHS, reaches a plateau. 
As expected, improvements in detection accuracy obtained by increasing $N_t$ come at the cost of \emph{higher} computational burden (cf. Tabs.~\ref{tab: Complexity comparison} and \ref{tab: complexitypery}).
To quantify this trade-off, Tab.~\ref{tab:combined_complexity} reports the overall runtime associated with both the \say{design complexity} (i.e., the cost of jointly optimizing the fusion rule and the RHS) and the \say{operational complexity} (i.e., the cost of executing the digital fusion once the RHS matrix and WL vector are designed).
The design-time results (Tab.~\ref{tab:combined_complexity}.a) clearly show the rapidly increasing runtime of the bFuC approaches as $N_t$ grows, in contrast to the more moderate requirements of the eFuC and IS-based designs (not depending on $N_t$). 
Conversely, the \say{operational complexity} (Tab.~\ref{tab:combined_complexity}.b) highlights that the proposed WL filter-bank structure efficiently exploits implicit target localization while avoiding the exponential growth of the GLR-based fusion rule with the number of sensors (which, in the considered case, theoretically scales as $\propto 2^{15}$).

\renewcommand{\arraystretch}{1.7}
\begin{table}[t]
\centering
\caption{Runtime comparison of design and operational fusion complexities for the WSN setup reported in Fig.~\ref{fig:Pd_vs_gridsize}.}
\label{tab:combined_complexity}

\begin{subtable}{\linewidth}
\centering
\caption{\textbf{Design Complexity}}
\begin{tabular}{ccccc}
\hline
\textbf{Rule} & $N_t=4$ & $N_t=9$ & $N_t=16$ & $N_t=25$ \\
\hline
\rowcolor[HTML]{D6D6FF}eFuC\text{-}0 & \multicolumn{4}{c}{$4.7\,\text{s}$} \\
\rowcolor[HTML]{D6D6FF}eFuC\text{-}1 & \multicolumn{4}{c}{$4.6\,\text{s}$} \\
\rowcolor[HTML]{F0C8F0}bFuC\text{-}0 & $20.0\,\text{s}$ & $38.2\,\text{s}$ & $64.6\,\text{s}$ & $99.8\,\text{s}$ \\
\rowcolor[HTML]{F0C8F0}bFuC\text{-}1 & $20.5\,\text{s}$ & $40.5\,\text{s}$ & $68.2\,\text{s}$ & $102.5\,\text{s}$ \\
\rowcolor[HTML]{f8e8e8}IS & \multicolumn{4}{c}{$0.3\,\text{s}$} \\
\hline
\end{tabular}
\end{subtable}

\medskip

\begin{subtable}{\linewidth}
\centering
\caption{\textbf{Operational Complexity}}
\begin{tabular}{ccccc}
\hline
\textbf{Rule} & $N_t=4$ & $N_t=9$ & $N_t=16$ & $N_t=25$ \\
\hline
\rowcolor[HTML]{DCDCDC}GLR & $312.4\,\text{s}$ & $336.4\,\text{s}$ & $407.8\,\text{s}$ & $483.1\,\text{s}$ \\
\rowcolor[HTML]{D6D6FF}eFuC & \multicolumn{4}{c}{$7.6\,\text{s}$} \\
\rowcolor[HTML]{F0C8F0}bFuC & $7.8\,\text{s}$ & $7.8\,\text{s}$ & $7.9\,\text{s}$ & $8.0\,\text{s}$ \\
\rowcolor[HTML]{f8e8e8}IS & \multicolumn{4}{c}{$7.3\,\text{s}$} \\
\hline
\end{tabular}
\end{subtable}

\renewcommand{\arraystretch}{1}
\end{table}

\subsection{WSN Size Impact on Target Detection via Holographic FC}

\begin{figure*}[t]
\centering{}\includegraphics[width=0.9\linewidth]{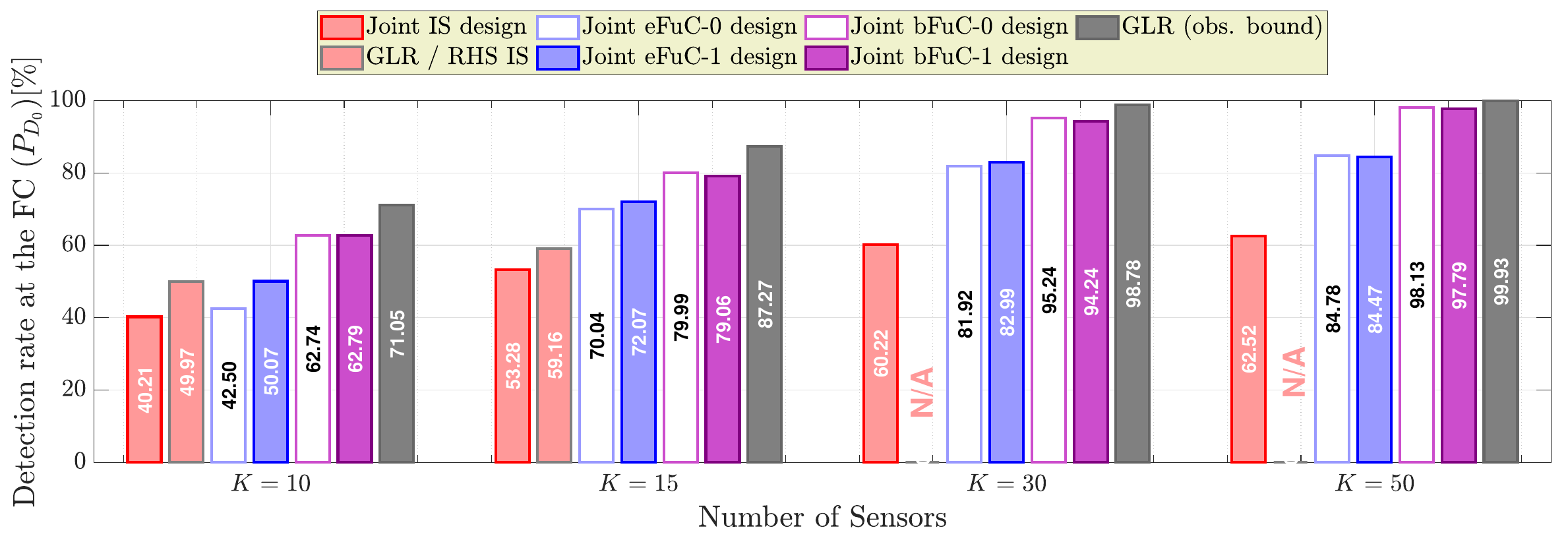}\caption{Impact of WSN size ($K$) on joint design in holographic DF via $P_{D_{0}}$ [\%], where the false-alarm rate is set to $P_{F_{0}}=10^{-2}$. A WSN with $K\in\{10,15,30,50\}$ sensors is considered.
Target strength is set to $\mathrm{SNR_{sen}}=15\,\mathrm{dB}$.
Holographic DF is implemented with an RHS made of $M=64$ elements and $N=1$ receive feed; channel noise variance is $\sigma_{w}^{2}=-50$ dBm. ``N/A'' indicates that GLR performance could not be computed due to prohibitive computational requirements.\label{fig:barchart_nosensors}}
\end{figure*}

Fig.~\ref{fig:barchart_nosensors} investigates the scalability of the proposed holographic DF architecture with respect to the number of sensors $K$, by reporting the detection probability $P_{D_0}$ at a fixed false-alarm rate $P_{F_0} = 0.01$ for four WSN configurations: $K \in \{10,15, 30, 50\}$.
As expected, the GLR observation bound also improves with increasing $K$ (cf. Eq.~\eqref{eq:GLRobs_RHS}), reflecting the fundamental benefit of higher sensing diversity.

The proposed DF approaches \emph{effectively exploit this diversity}. For instance, the eFuC-0 design improves its detection probability from approximately $71\%$ to $84\%$ as the network scales from $K=15$ to $K=50$ sensors. 
Even larger gains are observed for bFuC schemes (e.g., bFuC-0 achieves an $+18\%$ gain over the same range), highlighting the added value of filter-bank processing.
Interestingly, the performance gap between the bFuC and eFuC strategies widens with $K$, a trend that underscores the superior ability of the former to resolve the underlying composite hypothesis structure in large-scale WSNs.
Comparing with the same design strategy, it is apparent that for larger-scale WSNs solutions based on normal deflection (i.e. eFuC-0 and bFuC-0) are performing better and should be preferred because of their lower design complexity (in the case of bFuC-0, see Tab.~\ref{tab: Complexity comparison}).  
In contrast, the detection gains achieved by the joint IS design scale only marginally with $K$, thereby emphasizing the effectiveness of the proposed fusion strategies within the considered WSN size regime.
Finally, when using a GLR fusion rule with IS design on RHS matrix, performance is comparable to eFuC-1 design strategy for a small scale WSN (i.e. $K=10$). Conversely, performance gap widens at $K=15$.

\subsection{Performance Trends with Holographic FC parameters}

\begin{figure}
\centering{}\includegraphics[width=0.5\textwidth]{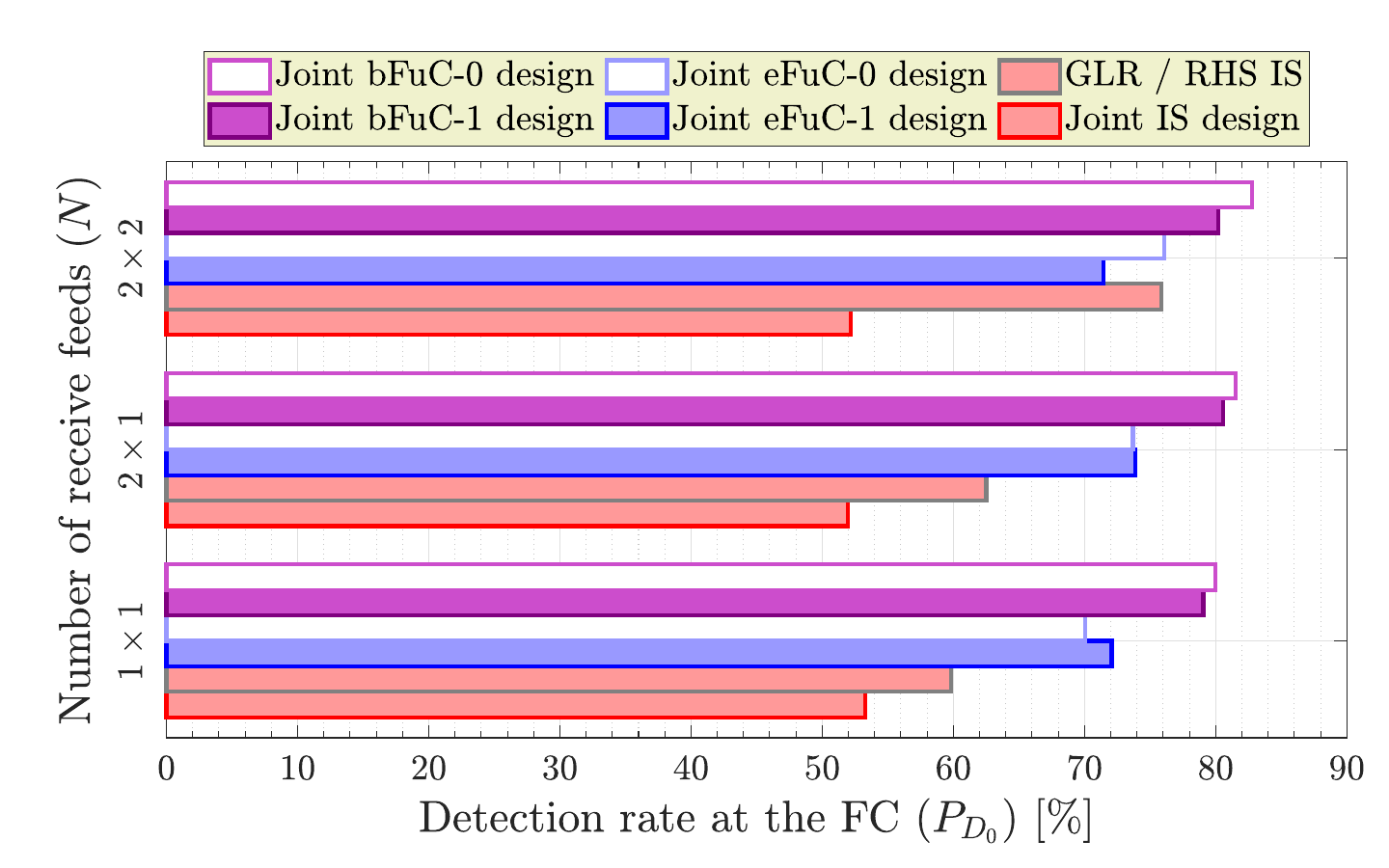}\caption{Impact of the number of receive feeds ($N$) on joint design in holographic DF via $P_{D_{0}}$ [\%], where the false-alarm rate is set to $P_{F_{0}}=10^{-2}$. A WSN with $K=15$ sensors is considered and target emission strength is set to $\mathrm{SNR_{sen}}=15\,\mathrm{dB}$. Holographic DF is implemented with an RHS made of $M=64$ elements, while three different feed configurations are considered: $N=\{1,2,4\}$ ($1\times1$, $2\times1$ and $2\times2$, respectively); channel noise variance is set to $\sigma_{w}^{2}=-50$ dBm.\label{fig:barchart_nofeeds}}
\end{figure}

Fig.~\ref{fig:barchart_nofeeds} examines the impact of increasing the number of receive feeds $N$ in the proposed holographic DF architecture by using the proposed joint design approaches. 
The analysis reports the detection probability $P_{D_0}$ at a fixed false-alarm rate $P_{F_0} = 0.01$ for three configurations: $N \in \{1, 2, 4\}$, corresponding to $1 \times 1$, $2 \times 1$ (linear), and $2 \times 2$ (planar) arrangements of the feed network.

Results show that the joint IS design \emph{is unable to capitalize on increased spatial sampling}, as performance remains virtually unchanged with growing $N$. This highlights a fundamental limitation in its capacity to exploit the additional degrees of freedom offered by multiple receive feeds.
Conversely, both eFuC-1 and bFuC-1 approaches exhibit significant performance improvements when increasing $N$ from $1$ to $2$--on top of their already higher baseline performance at $N = 1$. However, their gains plateau for $N = 4$, suggesting partial saturation in their ability to leverage further spatial diversity.

In contrast, eFuC-0 and bFuC-0 continue to benefit from larger feed arrays. Specifically, eFuC-0 improves its detection rate from approximately $70\%$ to $76\%$, while bFuC-0 sees a gain from $80\%$ to $83\%$ as $N$ increases from $1$ to $4$, thus demonstrating the appeal of using design solutions based on maximization of normal deflection when the number of feeds can grow.
These trends underscore the superior scalability of the proposed fusion mechanisms in leveraging feed diversity to enhance target detection performance.
Last, the GLR fusion combined with an IS-designed RHS matrix attains performance comparable to the eFuC solutions when $N=4$, underscoring the benefit of a more informative received vector.

\begin{figure*}[ht]
    \centering
    
    \begin{subfigure}[b]{0.45\textwidth}
       \centering{}\includegraphics[width=1\columnwidth]{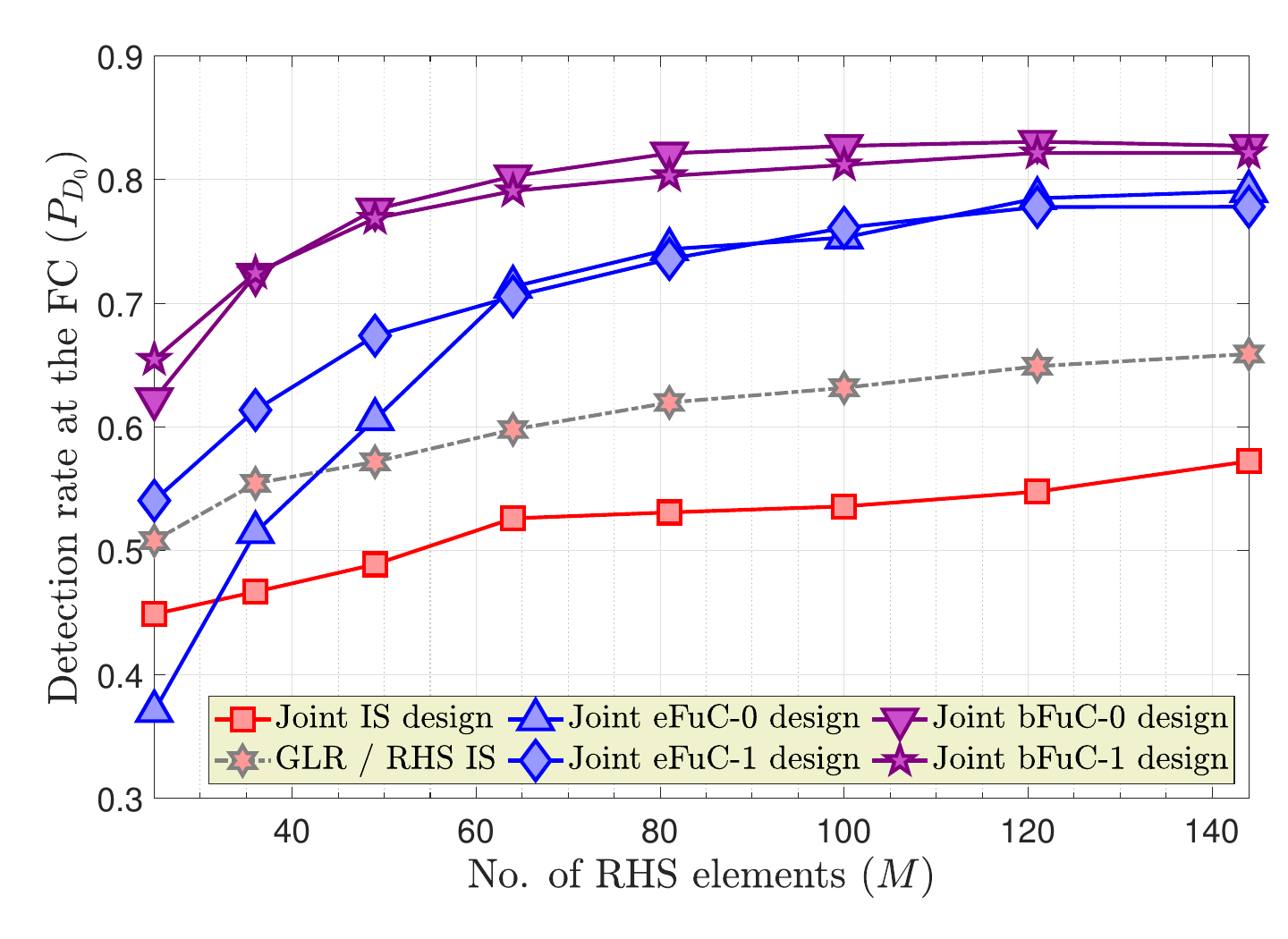}\caption{$N=1$ receive feed is considered.\label{fig:Pd0_vs_M_N1}}
    \end{subfigure}
    \hfill
    \begin{subfigure}[b]{0.45\textwidth}
       \centering{}\includegraphics[width=1\columnwidth]{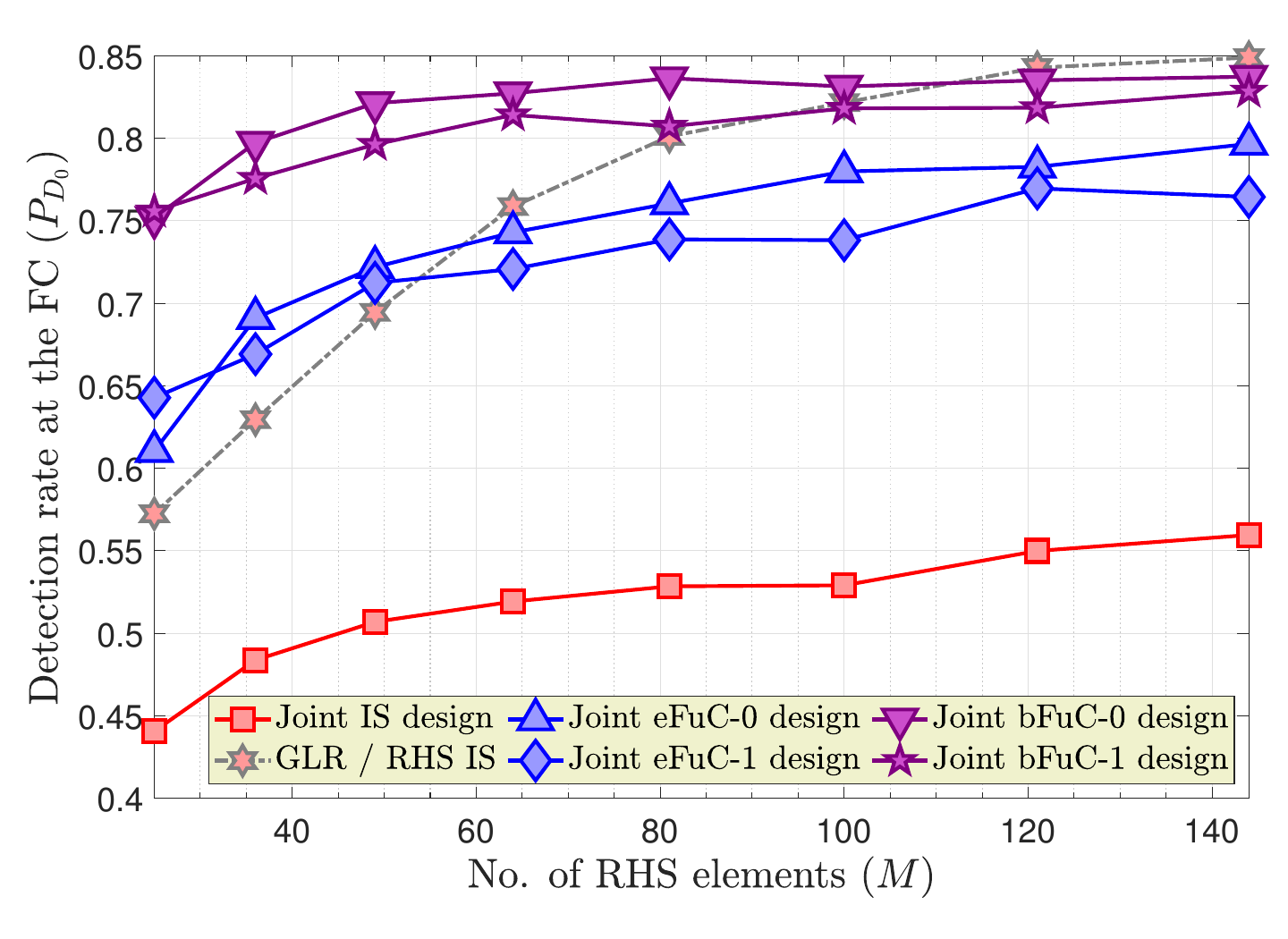}\caption{$N=4$ receive feeds are considered. \label{fig:Pd0_vs_M_N4}}
    \end{subfigure}
    
    \caption{$P_{D_{0}}$ vs number of RHS elements $M$ (with $P_{F_{0}}=0.01$) for two different feed scenarios ($N\in\{1,4\}$). WSN with $K=15$ sensors and target emission strength set to $\mathrm{SNR_{sen}}=15\,\mathrm{dB}$. Channel noise variance is set to $\sigma_{w}^{2}=-50$ dBm.}
    \label{fig:PD0_vs_M_combined}
\end{figure*}

Fig.~\ref{fig:PD0_vs_M_combined} analyzes the detection performance $P_{D_{0}}$ at a fixed false-alarm rate $P_{F_{0}} = 0.01$ as a function of the number of RHS elements $M$, with the aim of assessing the benefits of scaling the holographic aperture in the presence of a target at unknown position.
Across all design strategies, increasing $M$ consistently \emph{yields improved detection performance}, confirming the central role of aperture size in enhancing the effective sensing resolution. However, the rate and extent of (asymptotic) performance gain strongly depend on the adopted design strategy.

In the large-aperture regime ($M \geq 60$), performance tends to saturate, and the advantage of increasing the number of receive feeds (from $N = 1$ to $N = 4$) becomes negligible, compare Figs.~\ref{fig:Pd0_vs_M_N1}~and~\ref{fig:Pd0_vs_M_N4}.
In contrast, in the small- to medium-aperture regime ($M \in (25, 49)$), a larger number of feeds delivers measurable gains, particularly in scenarios where spatial sampling is inherently limited by aperture size.
The bFuC design family achieves superior performance even with small RHS sizes, effectively extracting target information from compact apertures. In comparison, eFuC strategies approach similar performance only in the large-aperture regime, reflecting a stronger dependence on aperture resolution.
Comparing eFuC-0 and eFuC-1 solutions, it is apparent that the choice of eFuC-0 is beneficial when the number of RHS elements is large or when the number of RF feeds grow.
A similar reasoning applies to the comparison between bFuC-0 and bFuC-1, with lower performance gap on the range considered.\\
On the contrary, the joint IS design (being sensor-agnostic and task-unaware) partly fails to leverage aperture scaling and exhibits a persistent detection gap, even asymptotically, relative to the proposed joint design techniques.
Last but not least, the GLR fusion with an IS-designed RHS only approaches the performance of the proposed methods under very \emph{demanding condition}s: specifically, when $N = 4$ (which significantly increases power consumption) and with a large RHS aperture of $M \ge 64$.
%

\section{Conclusions and Future Directions}
\label{Conclusions}
This work introduced the use of RHS to support channel-aware, DD of a non-cooperative target or PoI in a MIMO wireless sensor network. A holographic DF framework was proposed to jointly design the analog (RHS) and digital (fusion rule) components.
Given the lack of closed-form solutions for the GLR fusion rule and its computational burden, we developed four practical strategies--eFuC-0, eFuC-1, bFuC-0, and bFuC-1--based on two core principles. 
The first set of methods
(eFuC-0 and eFuC-1)
relies on second-order statistics of the received signals, which depend on the target’s position. This issue is resolved by substituting the unknown local detection probabilities with their expected values.
The second design (bFuC-0 and bFuC-1) adopts a filter-bank approach that approximates GLR-based fusion by optimizing the RHS configuration to enhance the average deflection of the filter outputs. The resulting optimization problems are non-convex but can be efficiently tackled using AO and MM techniques, leading to a simple iterative algorithm with closed-form updates.
Our results show that the proposed methods scale well with both the number of sensors ($K$) and the RHS aperture size (increasing $M$), demonstrating their suitability for large-scale deployments.

Looking forward, \emph{future research} will aim to broaden this framework to more advanced scenarios. This includes the integration of dynamic or stacked metasurfaces, direct RHS design under hardware constraints (e.g., quantized phase shifts), and robustness to imperfect channel state information or long-term channel statistics. 
Equally important, more complex sensing models will be explored, 
including soft-decision techniques at the local sensors~\cite{perez2025waveforms}, pushing the limits of the current holographic architecture.
Last but not least, further enhancements of the design methodology will be explored, including worst-case deflection formulations, strategies that exploit adaptive or data-driven grids and priors, alternative optimization frameworks beyond AO, and more informative initialization schemes.

\appendix
Leveraging convexity of $\ensuremath{\underbar{\ensuremath{\bm{\theta}}}^{\dagger}\,\bar{\bm{\Xi}}\,\underbar{\ensuremath{\bm{\theta}}}\,/\,c}$ in $\{\bm{\theta},c\}$, the objective function $g_{\mathrm{eFuC},i}(\bm{\theta})$ from Eq.~\eqref{eq: D_mFuC_i} can be minorized as~\cite{Yu2019}:
\begin{gather}
g_{\mathrm{eFuC},i}(\bm{\theta})=\frac{\underbar{\ensuremath{\bm{\theta}}}^{\dagger}\,\bar{\bm{\Xi}}\,\underbar{\ensuremath{\bm{\theta}}}}{\underbar{\ensuremath{\bm{\theta}}}^{\dagger}\,\bar{\bm{\Psi}}_{i}\,\underbar{\ensuremath{\bm{\theta}}}}\geq2\frac{\Re\left\{ \left(\underbar{\ensuremath{\bm{\theta}}}_{(\ell)}^{\star}\right)^{\dagger}\,\bar{\bm{\Xi}}\,\underbar{\ensuremath{\bm{\theta}}}\right\} }{\left(\underbar{\ensuremath{\bm{\theta}}}_{(\ell)}^{\star}\right)^{\dagger}\bar{\bm{\Psi}}_{i}\,\underbar{\ensuremath{\bm{\theta}}}_{(\ell)}^{\star}}\nonumber \\
-\frac{\left(\underbar{\ensuremath{\bm{\theta}}}_{(\ell)}^{\star}\right)^{\dagger}\,\bar{\bm{\Xi}}\,\underbar{\ensuremath{\bm{\theta}}}_{(\ell)}^{\star}}{\left(\left(\underbar{\ensuremath{\bm{\theta}}}_{(\ell)}^{\star}\right)^{\dagger}\bar{\bm{\Psi}}_{i}\,\underbar{\ensuremath{\bm{\theta}}}_{(\ell)}^{\star}\right)^{2}}\underbar{\ensuremath{\bm{\theta}}}^{\dagger}\,\bar{\bm{\Psi}}_{i}\,\underbar{\ensuremath{\bm{\theta}}}+\mathrm{const}\geq\nonumber \\
f_{\mathrm{eFuC},i}(\bm{\theta}|\bm{\theta}_{(\ell)}^{\star})+\left[g_{\mathrm{eFuC},i}(\underbar{\ensuremath{\bm{\theta}}}_{(\ell)}^{\star})-f_{\mathrm{eFuC},i}(\bm{\theta}_{(\ell)}^{\star}|\bm{\theta}_{(\ell)}^{\star})\right],\label{eq: objective_minorization_appendix}
\end{gather}
where the function $f_{\mathrm{eFuC},i}(\bm{\theta}|\bm{\theta}_{(\ell)}^{\star})$
is defined as
\begin{align}
f_{\mathrm{eFuC},i}(\bm{\theta}|\bm{\theta}_{(\ell)}^{\star}) & \triangleq2\frac{\Re\left\{ \left(\underbar{\ensuremath{\bm{\theta}}}_{(\ell)}^{\star}\right)^{\dagger}\,\bar{\bm{\Xi}}\,\underbar{\ensuremath{\bm{\theta}}}\right\} }{\left(\underbar{\ensuremath{\bm{\theta}}}_{(\ell)}^{\star}\right)^{\dagger}\bar{\bm{\Psi}}_{i}\,\underbar{\ensuremath{\bm{\theta}}}_{(\ell)}^{\star}}-\label{eq: surrogate objective appendix}\\
 & -\frac{\left(\underbar{\ensuremath{\bm{\theta}}}_{(\ell)}^{\star}\right)^{\dagger}\,\bar{\bm{\Xi}}\,\underbar{\ensuremath{\bm{\theta}}}_{(\ell)}^{\star}}{\left(\left(\underbar{\ensuremath{\bm{\theta}}}_{(\ell)}^{\star}\right)^{\dagger}\bar{\bm{\Psi}}_{i}\,\underbar{\ensuremath{\bm{\theta}}}_{(\ell)}^{\star}\right)^{2}}\left\{ \underbar{\ensuremath{\bm{\theta}}}^{\dagger}\,\lambda_{\mathrm{max}}\left(\bar{\bm{\Psi}}_{i}\right)\,\underbar{\ensuremath{\bm{\theta}}}\right.\nonumber \\
 & +2\Re\left\{ \left(\underbar{\ensuremath{\bm{\theta}}}_{(\ell)}^{\star}\right)^{\dagger}\left(\bar{\bm{\Psi}}_{i}\,-\,\lambda_{\mathrm{max}}\left(\bar{\bm{\Psi}}_{i}\,\right)\bm{I}_{2N}\right)\,\underbar{\ensuremath{\bm{\theta}}}\right\} \,.\nonumber 
\end{align}
Accordingly, the latter line of Eq.~\eqref{eq: objective_minorization_appendix} represents a reasonable surrogate objective to be maximized for MM. Also, since the term in the square brackets of Eq.~\eqref{eq: objective_minorization_appendix} does not depend on $\underbar{\ensuremath{\bm{\theta}}}$, it simply suffices to optimize the function $f_{\mathrm{eFuC},i}(\bm{\theta}|\bm{\theta}_{(\ell)}^{\star})$.
Then, by noticing that the second term of the objective in Eq.~\eqref{eq: surrogate objective appendix} does not depend on $\underbar{\ensuremath{\bm{\theta}}}$ (because of the constant modulus property of RHS shifts), optimization simply needs to be carried out on the first and third terms, namely:
\begin{gather}
\arg\max_{\bm{\theta}}f_{\mathrm{eFuC},i}(\bm{\theta}|\bm{\theta}_{(\ell)}^{\star})\triangleq\label{eq: surrogate objective appendix_2}\\
\arg\max_{\bm{\theta}}2\Re\left\{ \left[\frac{\left(\underbar{\ensuremath{\bm{\theta}}}_{(\ell)}^{\star}\right)^{\dagger}\,\bar{\bm{\Xi}}}{\left(\underbar{\ensuremath{\bm{\theta}}}_{(\ell)}^{\star}\right)^{\dagger}\bar{\bm{\Psi}}_{i}\,\underbar{\ensuremath{\bm{\theta}}}_{(\ell)}^{\star}}-\right.\right.\\
\left.\left.\frac{\left(\underbar{\ensuremath{\bm{\theta}}}_{(\ell)}^{\star}\right)^{\dagger}\,\bar{\bm{\Xi}}\,\underbar{\ensuremath{\bm{\theta}}}_{(\ell)}^{\star}}{\left(\left(\underbar{\ensuremath{\bm{\theta}}}_{(\ell)}^{\star}\right)^{\dagger}\bar{\bm{\Psi}}_{i}\,\underbar{\ensuremath{\bm{\theta}}}_{(\ell)}^{\star}\right)^{2}}\left(\underbar{\ensuremath{\bm{\theta}}}_{(\ell)}^{\star}\right)^{\dagger}\left(\bar{\bm{\Psi}}_{i}-\,\lambda_{\mathrm{max}}\left(\bar{\bm{\Psi}}_{i}\right)\bm{I}_{2M}\right)\right]\,\underbar{\ensuremath{\bm{\theta}}}\right\} \,.\nonumber 
\end{gather}
The latter optimization is in the form $\arg\max_{\ensuremath{\bm{\theta}}}\Re\left\{ \underline{\bm{\phi}}\ensuremath{^{\dagger}}\underline{\bm{\theta}}\right\} $
where $\bm{\phi}$ is a known complex vector. Accordingly, the solution is in closed form and equal to $\angle\text{\ensuremath{\underbar{\ensuremath{\bm{\theta}}}}}^{\star}=\angle\bm{\phi}$.

Differently, for bFuC-based design, it simply needs to be noticed that the objective $g_{\mathrm{bFuC},i}(\bm{\theta})=\sum_{j=1}^{N_{\mathrm{t}}}\frac{\underbar{\ensuremath{\bm{\theta}}}^{\dagger}\,\bm{\Xi}_{j}\,\underbar{\ensuremath{\bm{\theta}}}}{\underbar{\ensuremath{\bm{\theta}}}^{\dagger}\,\bm{\Psi}_{ij}\,\underbar{\ensuremath{\bm{\theta}}}}$ can be minorized (exploiting additivity) as:
\begin{gather}
g_{\mathrm{bFuC},i}(\bm{\theta})\geq\sum_{j=1}^{N_{\mathrm{t}}}f_{\mathrm{eFuC},i}^{j}(\bm{\theta}|\bm{\theta}_{(\ell)}^{\star})+\\
\sum_{j=1}^{N_{t}}\left[g_{\mathrm{eFuC},i}^{j}(\underbar{\ensuremath{\bm{\theta}}}_{(\ell)}^{\star})-f_{\mathrm{eFuC},i}^{j}(\bm{\theta}_{(\ell)}^{\star}|\bm{\theta}_{(\ell)}^{\star})\right]\,,\nonumber 
\end{gather}
where $f_{\mathrm{eFuC},i}^{j}(\bm{\theta}|\bm{\theta}_{(\ell)}^{\star})$ has a similar expression as in Eq.~\eqref{eq: surrogate objective appendix}.
Similarly as the eFuC case, the optimization of the surrogate is equivalent to maximizing $\sum_{j=1}^{N_{\mathrm{t}}}f_{\mathrm{eFuC},i}^{j}(\bm{\theta}|\bm{\theta}_{(\ell)}^{\star})$.
Also, because of the real operator properties and the linear dependence within each $f_{\mathrm{eFuC},i}^{j}(\bm{\theta}|\bm{\theta}_{(\ell)}^{\star})$ w.r.t. $\bm{\theta}$, the optimization can be re-arranged in the compact form $\arg\max_{\bm{\theta}}\Re\left\{ \left(\sum_{j=1}^{N_{\mathrm{t}}}\underline{\bm{\phi}}_{j}\right)^{\dagger}\ensuremath{\underbar{\ensuremath{\bm{\theta}}}}\right\} $. The latter expression highlights the closed form solution in Eq.~\eqref{eq: RHS optimization (Step 2) fc}. This concludes the proof.

\bibliographystyle{IEEEtran}
\bibliography{bibliography.bib}

@article{kay1998,
  title={Fundamentals of statistical signal processing, Vol. II: Detection Theory},
  author={S. M. Kay},
  journal={Signal Processing. Upper Saddle River, NJ: Prentice Hall},
  year={1998}
}

@article{quan2008,
  title={Optimal linear cooperation for spectrum sensing in cognitive radio networks},
  author={Z.~Quan and S.~Cui and A.~H.~Sayed},
  journal={{IEEE} J. Sel. Topics Signal Process.},
  volume={2},
  number={1},
  pages={28-40},
  month = feb,
  year={2008},
}

@article{Ciuonzo2012,
  title={Channel-aware decision fusion in distributed {MIMO} wireless sensor networks: Decode-and-fuse vs. decode-then-fuse},
  author={D.~Ciuonzo and G.~Romano and {P.~Salvo Rossi}},
  journal={{IEEE} Trans. Wireless Commun.},
  volume={11},
  number={8},
  pages={2976-2985},
  month = aug,
  year={2012},
}

@article{ciuonzo2015,
  title={Massive {MIMO} channel-aware decision fusion},
  author={D.~Ciuonzo and {P.~Salvo Rossi} and S.~Dey},
  journal={{IEEE} Trans. Signal Process.},
  volume={63},
  number={3},
  pages={604-619},
  mon = feb,
  year={2015},
}

@article{chawla2019,
  title={Distributed Detection in Massive {MIMO} Wireless Sensor Networks Under Perfect and Imperfect {CSI}},
  author={A. Chawla and A. Patel and A. K. Jagannatham and P. K. Varshney},
  journal={{IEEE} Trans. Signal Process.},
  volume={67},
  number={15},
  pages={4055-4068},
  year={2019},
  xxxpublisher={IEEE}
}

@article{sun2016majorization,
  title={Majorization-minimization algorithms in signal processing, communications, and machine learning},
  author={Sun, Ying and Babu, Prabhu and Palomar, Daniel P},
  journal={{IEEE} Trans. Signal Process.},
  volume={65},
  number={3},
  pages={794--816},
  year={2016},
  publisher={IEEE}
}

@article{Ahmed2022,
  title={Joint transmit and reflective beamformer design for secure estimation in {IRS-aided WSNs}},
  author={Ahmed, Mohammad Faisal and Rajput, Kunwar Pritiraj and Venkategowda, Naveen KD and Mishra, Kumar Vijay and Jagannatham, Aditya K},
 xxxjournal={IEEE Signal Processing Letters},
  journal={IEEE Signal Process. Lett.},
  volume={29},
  pages={692--696},
  year={2022},
  xxxpublisher={IEEE}
}

@article{fang2021,
  title={Over-the-air computation via reconfigurable intelligent surface},
  author={Fang, Wenzhi and Jiang, Yuning and Shi, Yuanming and Zhou, Yong and Chen, Wei and Letaief, Khaled B},
  journal={IEEE Trans. Commun.},
  volume={69},
  number={12},
  pages={8612--8626},
  year={2021},
  publisher={IEEE}
}

@article{picinbono1995,
  title={On deflection as a performance criterion in detection},
  author={Picinbono, Bernard},
  journal={IEEE Trans. Aerosp. Electron. Syst.},
  volume={31},
  number={3},
  pages={1072--1081},
  year={1995},
  xxxpublisher={IEEE}
}

@Article{Chen2004,
  Title                    = {Channel aware decision fusion in wireless sensor networks},
  Author                   = {B. Chen and R. Jiang and T. Kasetkasem and P. K. Varshney},
  Journal                  = {{IEEE} Trans. Signal Process.},
  Year                     = {2004},

  Month                    = dec,
  Number                   = {12},
  Pages                    = {3454-3458},
  Volume                   = {52},

  Doi                      = {10.1109/TSP.2004.837404}
}

@article{viswanathan1997,
  title={Distributed detection with multiple sensors Part I. Fundamentals},
  author={Viswanathan, Ramanarayanan and Varshney, Pramod K},
  xxxxjournal={Proceedings of the IEEE},
  journal={Proc. IEEE},
  volume={85},
  number={1},
  pages={54--63},
  year={1997},
  publisher={IEEE}
}

@article{Chawla2021,
  title={Distributed Detection for Centralized and Decentralized Millimeter Wave Massive {MIMO} Sensor Networks},
  author={Chawla, Apoorva and Singh, Rakesh Kumar and Patel, Adarsh and Jagannatham, Aditya K and Hanzo, Lajos},
  journal={IEEE Trans. Veh. Technol.},
  volume={70},
  number={8},
  pages={7665--7680},
  year={2021},
  publisher={IEEE}
}

@article{chair1986,
  title={Optimal data fusion in multiple sensor detection systems},
  author={Chair, Z and Varshney, PK},
  journal={IEEE Trans. Aerosp. Electron. Syst.},
  number={1},
  pages={98--101},
  year={1986},
  publisher={IEEE}
}

@article{reibman1987,
  title={Optimal detection and performance of distributed sensor systems},
  author={Reibman, Amy R and Nolte, LW},
  journal={IEEE Trans. Aerosp. Electron. Syst.},
  number={1},
  pages={24--30},
  year={1987},
  publisher={IEEE}
}

@article{zhang2008,
  title={Optimal power allocation for distributed detection over {MIMO} channels in wireless sensor networks},
  author={Zhang, Xin and Poor, H Vincent and Chiang, Mung},
  journal={{IEEE} Trans. Signal Process.},
  volume={56},
  number={9},
  pages={4124--4140},
  year={2008},
  publisher={IEEE}
}

@article{direnzo2020,
  title={Smart radio environments empowered by reconfigurable intelligent surfaces: How it works, state of research, and the road ahead},
  author={Di Renzo, Marco and Zappone, Alessio and Debbah, Merouane and Alouini, Mohamed-Slim and Yuen, Chau and De Rosny, Julien and Tretyakov, Sergei},
  journal={{IEEE} J. Sel. Areas Commun.},
  volume={38},
  number={11},
  pages={2450--2525},
  year={2020},
  publisher={IEEE}
}

@inproceedings{mudkey2022wireless,
  title={Wireless Inference Gets Smarter: {RIS-}assisted Channel-Aware {MIMO} Decision Fusion},
  author={Mudkey, Nishanth and Ciuonzo, Domenico and Zappone, Alessio and Salvo Rossi, Pierluigi},
  booktitle={IEEE 12th Sensor Array and Multichannel Signal Processing Workshop (SAM)},
  pages={26--30},
  year={2022},
  xxxxorganization={IEEE}
}

@article{ge2024ris,
  title={{RIS-Assisted} Cooperative Spectrum Sensing for Cognitive Radio Networks},
  author={Ge, Jungang and Liang, Ying-Chang and Wang, Shuo and Sun, Chen},
  xxxjournal={IEEE Transactions on Wireless Communications},
  journal={{IEEE} Trans. Wireless Commun.},
  year={2024},
  volume={23},
  number={9},
  pages={12547-12562},
  publisher={IEEE}
}

@inproceedings{rajput2024joint,
  title={Joint Transmit Precoders and Passive Reflection Beamformer Design in {IRS-Aided IoT} Networks},
  author={Rajput, Kunwar Pritiraj and Wu, Linlong and Shankar, MR Bhavani and Varshney, Pramod K},
  booktitle={IEEE International Conference on Acoustics, Speech and Signal Processing (ICASSP)},
  pages={156--160},
  year={2024},
  xxxorganization={IEEE}
}

@article{zhang2022worst,
  title={Worst-case design for {RIS-aided} over-the-air computation with imperfect {CSI}},
  author={Zhang, Wenhui and Xu, Jindan and Xu, Wei and You, Xiaohu and Fu, Weijie},
  journal={{IEEE} Commun. Lett.},
 xxxjournal={IEEE Communications Letters},
  volume={26},
  number={9},
  pages={2136--2140},
  year={2022},
  xxxpublisher={IEEE}
}

@article{zhai2022beamforming,
  title={Beamforming design based on two-stage stochastic optimization for {RIS-assisted} over-the-air computation systems},
  author={Zhai, Xiongfei and Han, Guojun and Cai, Yunlong and Hanzo, Lajos},
  xxxjournal={IEEE Internet of Things Journal},
  journal={IEEE Internet Things J.},
   year={2022},
  volume={9},
  number={7},
  pages={5474-5488},
  publisher={IEEE}
}

@inproceedings{zhao2023ris,
  title={{RIS-Assisted} {CSIT-free} Data Fusion with Timing Misalignment},
  author={Zhao, Yaqiong and Xu, Wei and Ye, Xinquan},
  booktitle={9th IEEE International Conference on Computer and Communication Engineering (ICCCE)},
  pages={12--17},
  year={2023},
  xxxorganization={IEEE}
}

@inproceedings{ellingson2021path,
  title={Path loss in reconfigurable intelligent surface-enabled channels},
  author={Ellingson, Steven W},
  booktitle={IEEE 32nd Annual International Symposium on Personal, Indoor and Mobile Radio Communications (PIMRC)},
  pages={829--835},
  year={2021},
  xxxorganization={IEEE}
}

@article{jamali2020intelligent,
  title={Intelligent surface-aided transmitter architectures for millimeter-wave ultra massive {MIMO} systems},
  author={Jamali, Vahid and Tulino, Antonia M and Fischer, Georg and M{\"u}ller, Ralf R and Schober, Robert},
  journal={IEEE Open Journal of the Communications Society},
  volume={2},
  pages={144--167},
  year={2020},
  publisher={IEEE}
}

@article{feng2023near,
  title={Near-field modelling and performance analysis for extremely large-scale {IRS} communications},
  author={Feng, Chao and Lu, Haiquan and Zeng, Yong and Li, Teng and Jin, Shi and Zhang, Rui},
  xxxjournal={IEEE Transactions on Wireless Communications},
  journal={{IEEE} Trans. Wireless Commun.},
  year={2023},
  publisher={IEEE}
}

@article{tang2020wireless,
  title={Wireless communications with reconfigurable intelligent surface: Path loss modeling and experimental measurement},
  author={Tang, Wankai and Chen, Ming Zheng and Chen, Xiangyu and Dai, Jun Yan and Han, Yu and Di Renzo, Marco and Zeng, Yong and Jin, Shi and Cheng, Qiang and Cui, Tie Jun},
  xxxjournal={IEEE Transactions on Wireless Communications},
  journal={{IEEE} Trans. Wireless Commun.},
  volume={20},
  number={1},
  pages={421--439},
  year={2020},
  publisher={IEEE}
}

@inproceedings{Ciuonzo2025icassp,
  title={Massive {MIMO} Channel-aware
Decision Fusion aided by Reconfigurable Intelligent Surfaces},
  author={Ciuonzo, Domenico and Zappone, Alessio and Di Renzo, Marco and Wu, Linlong},
  booktitle={IEEE International
Conference on Acoustics, Speech, and Signal Processing (ICASSP)},
 year = {2025}
 }

@article{li2007distributed,
  title={Distributed detection in wireless sensor networks using a multiple access channel},
  author={Li, Wenjun and Dai, Huaiyu},
  xxxjournal={IEEE Transactions on Signal Processing},
  journal={{IEEE} Trans. Signal Process.},
  volume={55},
  number={3},
  pages={822--833},
  year={2007},
  publisher={IEEE}
}

@article{zhang2008optimal,
  title={Optimal power allocation for distributed detection over {MIMO} channels in wireless sensor networks},
  author={Zhang, Xin and Poor, H Vincent and Chiang, Mung},
  xxxjournal={IEEE Transactions on Signal Processing},
  journal={{IEEE} Trans. Signal Process.},
  volume={56},
  number={9},
  pages={4124--4140},
  year={2008},
  publisher={IEEE}
}

@article{jiang2007multiuser,
  title={Multiuser {MIMO-OFDM} for next-generation wireless systems},
  author={Jiang, Ming and Hanzo, Lajos},
  xxxjournal={Proceedings of the IEEE},
  journal={Proc. IEEE},
  volume={95},
  number={7},
  pages={1430--1469},
  year={2007},
  publisher={IEEE}
}

@article{lu2014overview,
  title={An overview of massive {MIMO}: Benefits and challenges},
  author={Lu, Lu and Li, Geoffrey Ye and Swindlehurst, A Lee and Ashikhmin, Alexei and Zhang, Rui},
  xxxxjournal={IEEE journal of selected topics in signal processing},
  journal={IEEE J. Sel. Topics Signal Process.},
  volume={8},
  number={5},
  pages={742--758},
  year={2014},
  publisher={IEEE}
}

@article{jiang2015massive,
  title={Massive {MIMO} for wireless sensing with a coherent multiple access channel},
  author={Jiang, Feng and Chen, Jie and Swindlehurst, A Lee and L{\'o}pez-Salcedo, Jos{\'e} A},
  xxxjournal={IEEE Transactions on Signal Processing},
  journal={{IEEE} Trans. Signal Process.},
  volume={63},
  number={12},
  pages={3005--3017},
  year={2015},
  publisher={IEEE}
}

@article{huang2020holographic,
  title={Holographic {MIMO} surfaces for {6G} wireless networks: Opportunities, challenges, and trends},
  author={Huang, Chongwen and Hu, Sha and Alexandropoulos, George C and Zappone, Alessio and Yuen, Chau and Zhang, Rui and Di Renzo, Marco and Debbah, Merouane},
  xxxjournal={IEEE Wireless Communications},
  journal={IEEE Wireless Commun.},
  volume={27},
  number={5},
  pages={118--125},
  year={2020},
  publisher={IEEE}
}

@article{interdonato2024approaching,
  title={Approaching Massive {MIMO} Performance with Reconfigurable Intelligent Surfaces: We Do Not Need Many Antennas},
  author={Interdonato, Giovanni and Di Murro, Francesca and D’Andrea, Carmen and Di Gennaro, Giovanni and Buzzi, Stefano},
  xxxjournal={IEEE Transactions on Communications},
  journal={IEEE Trans. Commun.},
  year={2024},
  xxxpublisher={IEEE}
}

@ARTICLE{DegliEsposti2022,
  author={Degli-Esposti, Vittorio and Vitucci, Enrico M. and Renzo, Marco Di and Tretyakov, Sergei A.},
  xxxjournal={IEEE Transactions on Antennas and Propagation},
  journal={IEEE Trans. Antennas Propag.}, 
  title={Reradiation and Scattering From a Reconfigurable Intelligent Surface: A General Macroscopic Model}, 
  year={2022},
  volume={70},
  number={10},
  pages={8691-8706}
}

@article{abrardo2021intelligent,
  title={Intelligent reflecting surfaces: Sum-rate optimization based on statistical position information},
  author={Abrardo, Andrea and Dardari, Davide and Di Renzo, Marco},
  xxxjournal={IEEE Transactions on Communications},
  journal={IEEE Trans. Commun.},
  volume={69},
  number={10},
  pages={7121--7136},
  year={2021},
  publisher={IEEE}
}

@book{lehmann2022,
  title={Testing statistical hypotheses},
  author={Lehmann, Erich Leo and Romano, Joseph P},
  volume={4},
  year={2022},
  publisher={Springer}
}

@article{ciuonzo2017quantizer,
  title={Quantizer design for generalized locally optimum detectors in wireless sensor networks},
  author={Ciuonzo, Domenico and Salvo Rossi, Pierluigi },
  journal={IEEE Wireless Communications Letters},
  volume={7},
  number={2},
  pages={162--165},
  year={2017},
  publisher={IEEE}
}

@ARTICLE{Ciuonzo2025IoT,
  author={Ciuonzo, Domenico and Zappone, Alessio and Di Renzo, Marco},
  xxxjournal={IEEE Internet of Things Journal},
  journal={IEEE Internet Things J.},
  title={Channel-Aware Holographic Decision Fusion}, 
  year={2025},
  volume={},
  number={},
  pages={1-1},
  doi={10.1109/JIOT.2025.3574794}
}

@inproceedings{Yu2019,
  title={Enabling secure wireless communications via intelligent reflecting surfaces},
  author={Yu, Xianghao and Xu, Dongfang and Schober, Robert},
  booktitle={IEEE Global Communications Conference (GLOBECOM)},
  pages={1--6},
  year={2019},
  xxxorganization={IEEE}
}

@article{blum2002distributed,
  title={Distributed detection with multiple sensors II. Advanced topics},
  author={Blum, Rick S and Kassam, Saleem A and Poor, H Vincent},
  xxxjournal={Proceedings of the IEEE},
  journal={Proc. {IEEE}},
  volume={85},
  number={1},
  pages={64--79},
  year={2002},
  publisher={IEEE}
}

@Article{cheng2019multibit,
  author    = {X. Cheng and D. Ciuonzo and {P. Salvo Rossi}},
  journal   = {IEEE Trans. Aerosp. Electron. Syst.},
  title     = {Multibit Decentralized Detection Through Fusing Smart and Dumb Sensors Based on {Rao} Test},
  year      = {2019},
  number    = {2},
  pages     = {1391-1405},
  volume    = {56},
  publisher = {IEEE},
}

@article{yang2023hybrid,
  title={Hybrid quantized signal detection with a bandwidth-constrained distributed radar system},
  author={Yang, Shixing and Lai, Yangming and Jakobsson, Andreas and Yi, Wei},
  journal={{IEEE} Trans. Aerosp. Electron. Syst.},
  volume={59},
  number={6},
  pages={7835--7850},
  year={2023},
  publisher={IEEE}
}

@article{mao2024multi,
  title={Multi-bit distributed detection of sparse stochastic signals over error-prone reporting channels},
  author={Mao, Linlin and Yan, Shefeng and Sui, Zeping and Li, Hongbin},
  journal={IEEE Trans. on Signal and Information Processing over Networks},
  volume={10},
  pages={881--893},
  year={2024},
  publisher={IEEE}
}

@article{molisch2017hybrid,
  title={Hybrid beamforming for massive {MIMO}: A survey},
  author={Molisch, Andreas F and Ratnam, Vishnu V and Han, Shengqian and Li, Zheda and Nguyen, Sinh Le Hong and Li, Linsheng and Haneda, Katsuyuki},
  xxxjournal={IEEE Communications magazine},
  journal = {IEEE Commun. Mag.},
  volume={55},
  number={9},
  pages={134--141},
  year={2017},
  publisher={IEEE}
}

@article{willsky2003generalized,
  title={A generalized likelihood ratio approach to the detection and estimation of jumps in linear systems},
  author={Willsky, Alan and Jones, H},
  xxxjournal={IEEE Transactions on Automatic control},
  journal={IEEE Trans. Autom. Control},
  volume={21},
  number={1},
  pages={108--112},
  year={2003},
  publisher={IEEE}
}

@article{ciuonzo2017rician,
  title={Rician {MIMO} channel-and jamming-aware decision fusion},
  author={Ciuonzo, Domenico and Aubry, Augusto and Carotenuto, Vincenzo},
  xxxjournal={IEEE Transactions on Signal Processing},
  journal={{IEEE} Trans. Signal Process.},
  volume={65},
  number={15},
  pages={3866--3880},
  year={2017},
  publisher={IEEE}
}

@ARTICLE{Ciuonzo2025single,
  author={Ciuonzo, D.},
  xxxjournal={IEEE Signal Processing Letters}, 
  journal={IEEE Signal Process. Lett.},
  title={Score-Based Fading-Aware Decision Fusion}, 
  year={2025},
  volume={32},
  number={},
  pages={2952-2956}
}

@article{perez2025waveforms,
  title={Waveforms for Computing Over the Air: A groundbreaking approach that redefines data aggregation},
  author={P{\'e}rez-Neira, Ana and Martinez-Gost, Marc and {\c{S}}ahin, Alphan and Razavikia, Saeed and Fischione, Carlo and Huang, Kaibin},
  journal={IEEE Signal Process. Mag.},
  volume={42},
  number={2},
  pages={57--77},
  year={2025},
  xxxpublisher={IEEE}
}

\begin{IEEEbiography}[{\includegraphics[width=1in,height=1.25in,clip,keepaspectratio]{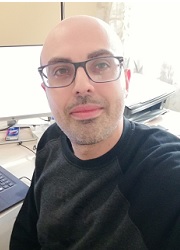}}]{Domenico~Ciuonzo} (Senior Member, IEEE)  is an Associate Professor at University of Naples \say{Federico II}. 
Since 2011, he has been holding several visiting researcher appointments (NATO CMRE, UConn, NTNU, CTTC). 
He is the recipient of two Best Paper awards (IEEE ICCCS 2019 and Elsevier Computer Networks 2020), the 2019 Exceptional Service award from \textsc{IEEE AESS}, the 2020 Early-Career Technical Achievement award from \textsc{IEEE Sensors Council} for sensor networks/systems and the 2021 Early-Career Award from \textsc{IEEE AESS}.
His research interests include data fusion, wireless sensor networks, the IoT, and machine learning.
\end{IEEEbiography}

\begin{IEEEbiography}[{\includegraphics[width=1in,height=1.25in,clip,keepaspectratio]{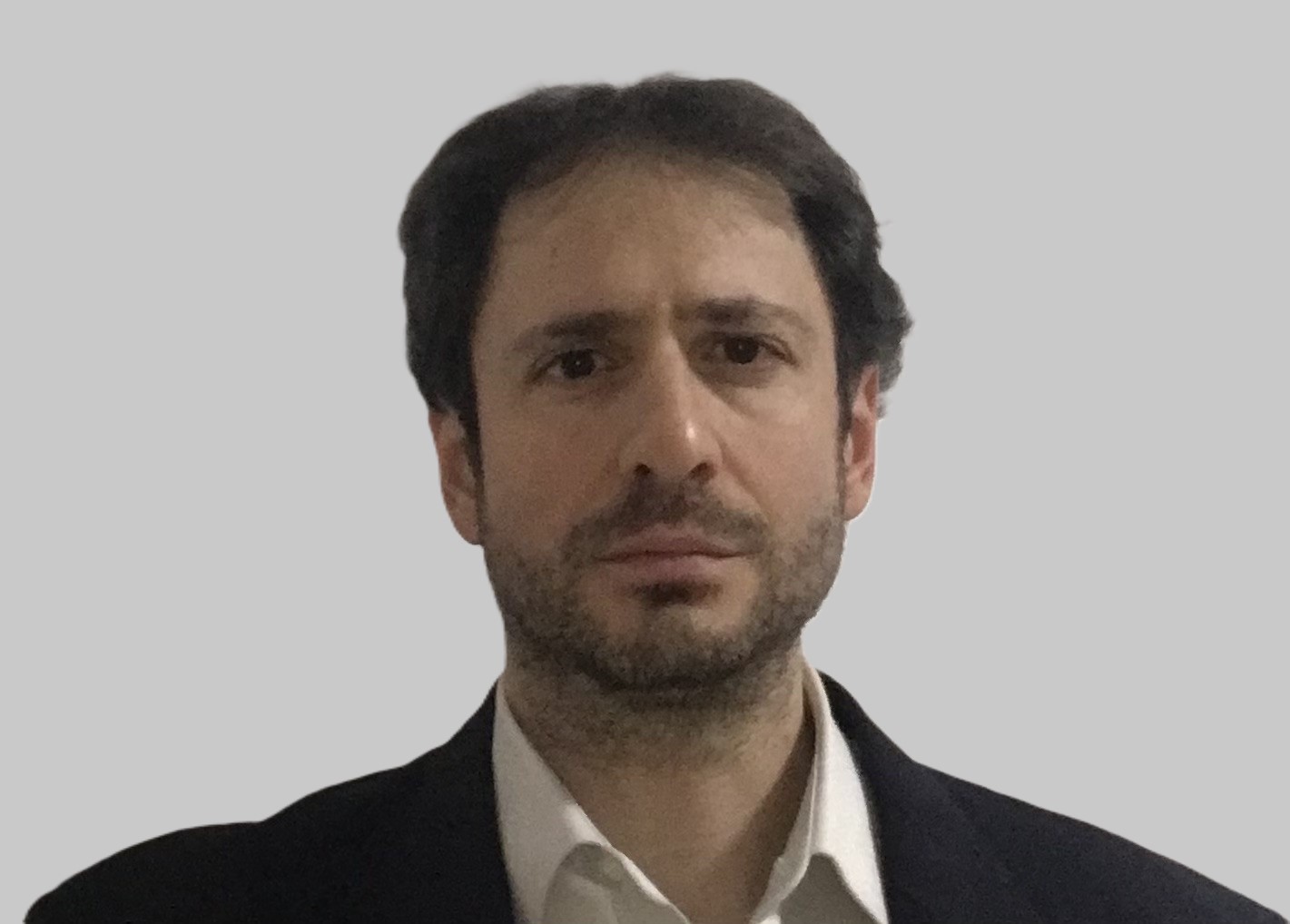}}]{Alessio Zappone} (Fellow, IEEE) is a tenured professor at the University of Cassino and Southern Lazio. His research interests lie in the area of communication theory and signal processing, with main focus on optimization techniques for resource allocation and energy efficiency maximization. For his research, Alessio received several award among which the IEEE Marconi Prize Paper Award in Wireless Communications in 2021, the IEEE Communications Society Fred W. Ellersick Prize in 2023, and the IEEE Communications Society Best Tutorial Paper Award in 2024. Alessio serves as editor of the \textsc{IEEE Transactions on Wireless Communications}, area editor of the \textsc{IEEE Communications Letters}, and has served as senior editor of the \textsc{IEEE Signal Processing Letters} and guest editor of two \textsc{IEEE Journal on Selected Areas in Communications} special issues.
\end{IEEEbiography}

\begin{IEEEbiography}[{\includegraphics[width=1in,height=1.25in,clip,keepaspectratio]{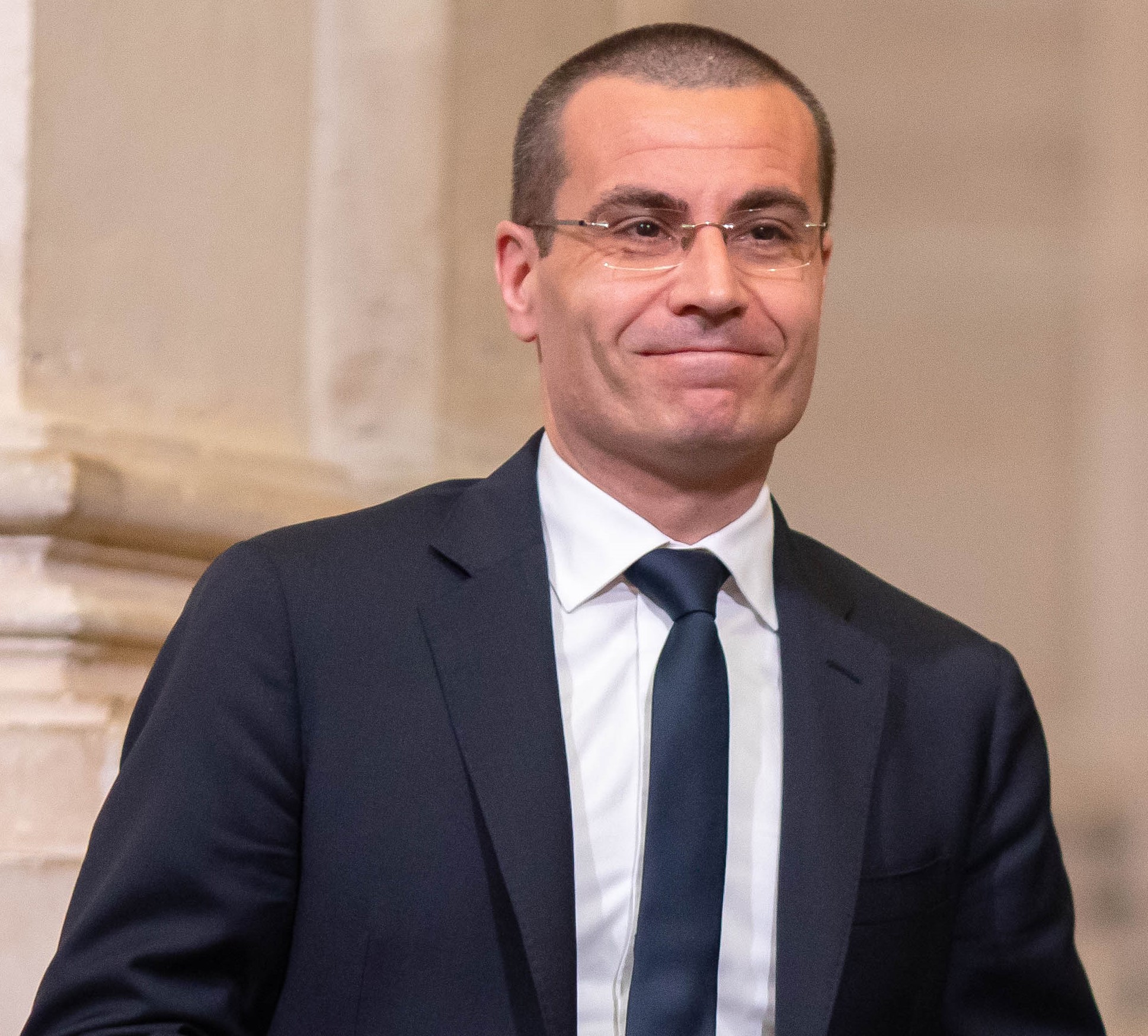}}]{Marco~Di~Renzo} (Fellow, IEEE) received the Laurea (cum laude) and Ph.D. degrees in electrical engineering from the University of L’Aquila, Italy, in 2003 and 2007, respectively, and the Habilitation à Diriger des Recherches (Doctor of Science) degree from University Paris-Sud (currently Paris-Saclay University), France, in 2013. Currently, he is a CNRS Research Director (Professor) and the Head of the Intelligent Physical Communications group with the Laboratory of Signals and Systems (L2S) at CNRS \& CentraleSupélec, Paris-Saclay University, Paris, France, as well as a Chair Professor in Telecommunications Engineering with the Centre for Telecommunications Research -- Department of Engineering, King’s College London, London, United Kingdom. He was a France-Nokia Chair of Excellence in ICT at the University of Oulu (Finland), a Tan Chin Tuan Exchange Fellow in Engineering at Nanyang Technological University (Singapore), a Fulbright Fellow at The City University of New York (USA), a Nokia Foundation Visiting Professor at Aalto University (Finland), and a Royal Academy of Engineering Distinguished Visiting Fellow at Queen’s University Belfast (U.K.). He is a Fellow of the IEEE, IET, EURASIP, and AAIA; an Academician of AIIA; an Ordinary Member of the European Academy of Sciences and Arts, an Ordinary Member of the Academia Europaea; an Ambassador of the European Association on Antennas and Propagation; and a Highly Cited Researcher. His recent research awards include the Michel Monpetit Prize conferred by the French Academy of Sciences, the IEEE Communications Society Heinrich Hertz Award, and the IEEE Communications Society Marconi Prize Paper Award in Wireless Communications. He served as the Editor-in-Chief of IEEE Communications Letters from 2019 to 2023. His current main roles within the IEEE Communications Society include serving as a Voting Member of the Fellow Evaluation Standing Committee, as the Chair of the Publications Misconduct Ad Hoc Committee, and as the Director of Journals. Also, he is on the Editorial Board of the Proceedings of the IEEE.
\end{IEEEbiography}

\begin{IEEEbiography}[{\includegraphics[width=1in,height=1.25in,clip,keepaspectratio]{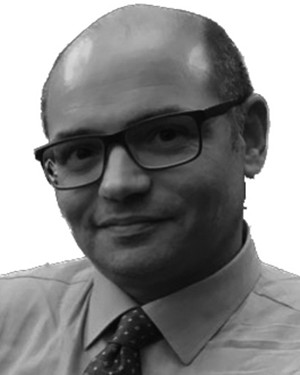}}]{Ciro d'Elia} received the M.S. degree (Hons.) in telecommunication engineering and the Ph.D. degree in computer science and electronic engineering from the University of Naples Federico II, Naples, Italy, in 1998 and 2001, respectively. Since October 2001, he has been a Researcher with the Department of Electrical and Information Engineering (DIEI), University of Cassino and Southern Lazio, Cassino, Italy, where he teaches the M.S. courses on Telecommunication Networks, Telematics, Image Processing and Transmission, and the Ph.D. course on Object-Oriented Design for Signal Processing. Since 2002, he has been the Technical Director of the Informatics and Telecommunication Laboratory, DIEI. He has participated in several research projects funded by the Italian Ministry of Education University and Research (MIUR), the Italian Space Agency (ASI), the German Aerospace Center (DLR), and the European Space Agency (ESA). He held contracts from several national companies: Telecom Italia, Sogin, MBDA Italia, Urmet, and ACEA ATO2 and foreign companies: MetaSensing, The Netherlands, VicomTech, Spain. His research interests are in statistical signal and image processing and mining, information extraction from remotely sensed data, telecommunication networks, telematics, network security, the IoT, and cybersecurity.

\end{IEEEbiography}

\end{document}